\documentclass[a4paper,12pt]{JHEP3}

\usepackage{amsmath,epsfig}
\usepackage{amssymb,amsfonts}
\usepackage{latexsym}
\usepackage[latin1]{inputenc}
\usepackage{xcolor}

\usepackage{graphicx}
\usepackage{longtable}

\relax
\renewcommand{\theequation}{\arabic{section}.\arabic{equation}}
\def\be{\begin{equation}}
\def\ee{\end{equation}}
\def\bea{\begin{eqnarray}}
\def\eea{\end{eqnarray}}

\newcommand\fverb{\setbox\pippobox=\hbox\bgroup\verb}
\newcommand\fverbdo{\egroup\medskip\noindent%
                        \fbox{\unhbox\pippobox}\ }
\newcommand\fverbit{\egroup\item[\fbox{\unhbox\pippobox}]}

\newcommand{\bear}{\begin{eqnarray}}

\newcommand{\eear}{\end{eqnarray}}

\newcommand{\bsea}{\begin{subeqnarray}}
\newcommand{\esea}{\end{subeqnarray}}
\newbox\pippobox

\def\6{\partial}

\def\a{\alpha}

\def\pa{\partial}

\def\m{\mu}
\def\n{\nu}
\def\r{\rho}
\def\s{\sigma}

\def\sp{\;\;\;,\;\;\;}

\def\sq
\def\a{\alpha}

\def\tr{{\rm Tr}}

\def\hri#1#2{\href{http://arxiv.org/abs/#1}{[ArXiv:#1]#2}}
\def\hre#1#2{\href{http://arxiv.org/abs/#1/#2}{[ArXiv:#1/#2]}}

\allowdisplaybreaks[1]

\title{Charge-hyperscaling violating Lifshitz hydrodynamics from black-holes.}
\author{\Large   Elias Kiritsis$^{a,b,c}$, Yoshinori Matsuo$^{a,b}$\\
~\\
~\\
$^a$\href{http://hep.physics.uoc.gr}{Crete Center for Theoretical Physics},
Department of Physics, University of Crete, 71003 Heraklion, Greece.
~\\
$^b$Crete Center for Quantum Complexity and Nanotechnology,
Department of Physics, University of Crete, 71003 Heraklion, Greece.
~\\
$^c$\href{http://www.apc.univ-paris7.fr/APC_CS/}{APC}, Astrparticule et Cosmologie, Universit\'e Paris Diderot, CNRS/IN2P3, CEA/Irfu, Observatoire de Paris, Sorbonne Paris Cit\'e, 10, rue Alice Domon et L\'eonie Duquet, 75205 Paris Cedex 13, France.
~\\
E-mail: \href{http://hep.physics.uoc.gr/~kiritsis/}{http://hep.physics.uoc.gr/~kiritsis/},
matsuo@physics.uoc.gr
}

\preprint{CCTP-2015-14\\
CCQCN-2015-94}

\abstract{Non-equilibrium black hole horizons are considered in scaling theories with generic Lifshitz invariance and an unbroken U(1) symmetry.
There is also  charge-hyperscaling violation associated with a non-trivial conduction exponent. The boundary stress tensor is computed and renormalized and the associated hydrodynamic equations derived. Upon a non-trivial redefinition of boundary sources associated with the U(1) gauge field, the equations are mapped to the standard non-relativistic hydrodynamics equations coupled to a mass current and an external Newton potential in accordance with the general theory of \cite{Hartong:2015wxa}.
The shear viscosity to entropy ratio is the same as in the relativistic case.
}


\begin{document}

\section{Introduction}\label{intro}

The AdS/CFT correspondence
\cite{Maldacena:1997re, Gubser:1998bc, Witten:1998qj, Witten:1998zw}
relates the anti-de Sitter (AdS) space-time to
conformal field theory (CFT) on the boundary.
It gives a semiclassical description of strong coupling physics
in the dual field theory, in terms of string theory or its low-energy limit: (super) gravity.
At finite temperature and in the long wavelength regime,
the dual field theory can be effectively described by fluid mechanics
and it can be related to black holes in AdS space-time.

The fluid/gravity correspondence was
first studied using linear response theory
\cite{Son:2002sd, Policastro:2002se, Policastro:2002tn, Baier:2007ix}.
Subsequently, fully dynamic descriptions were
studied using boosted black holes in asymptotically AdS geometries
that led to relativistic fluid dynamics in the dual CFT \cite{Bhattacharyya:2008jc}\footnote{A related work was presented in \cite{{Baier:2007ix}}.}.
In this formalism, the fluid variables are encoded in the near-equilibrium black hole solution  and the
fluid equations appear  as constraints on the solution imposed by the bulk equations of motion.

Recently, generalizations of the AdS/CFT correspondence to
theories with non-relativistic scaling symmetry have been studied.
In particular, many condensed matter systems have critical points with
non-relativistic scale invariance
\cite{Eagles:1969zz, Nozieres:1985zz, O'Hara:2002zz, Regal:2004zza, Bartenstein:2004zza}.
Some of these systems have Lifshitz  or Schr\"odinger symmetry
\cite{Hornreich:1975zz, Mehen:1999nd, Ardonne:2003wa}.
Moreover, the hydrodynamics of charge and energy in such systems may be interesting as has been argued recently for the case of cold fermions at unitarity, \cite{fu}, other strongly correlated systems, \cite{zaa} and  graphene, \cite{gra}.

Holographic techniques have been  generalized to geometries
with Lifshitz or Schr\"odinger symmetry in connection with  applications
to condensed matter systems
\cite{Son:2008ye, Balasubramanian:2008dm, kachru, taylor, Guica:2010sw, cgkkm, gk1}.
In particular, in \cite{cgkkm, gk1, gk2}, all quantum critical holographic scaling theories with a $U(1)$ symmetry respecting translation invariance and spatial rotation invariance were classified in terms of three scaling exponents.
Two of them $(z,\theta)$ appear in the metric while another exponent, $\psi$
appears in the profile of  the $U(1)$ gauge field (it is referred to as $\zeta$ in \cite{cgkkm, gk1, gk2}).%
\footnote{This charge exponent controls the anomalous scaling of the charge density, even though charge is conserved.It  has been also introduced independently in \cite{hartong-obers} and was studied in more detail in \cite{G1} and \cite{karch}. The reason for the existence of an anomalous charge exponent despite conservation is the RG running of the bulk coupling for charged degrees of freedom.}
The exponent $z$ is the Lifshitz (dynamical) scaling exponent and
$\theta$ is the hyperscaling-violation exponent, \cite{gk1,sh}.
Even though such theories have been studied intensively, many  aspects are still unclear and in particular, hydrodynamics with Lifshitz scaling symmetry is not fully understood.

More recently, it was found that the boundary theory dual to space-times  with Lifshitz asymptotics
can be described in terms of the torsional Newton-Cartan gravity theory,
which is a novel extension of the Newton-Cartan gravity with a specific torsion tensor.
The application of the Newton-Cartan theory to non-relativistic  condensed matter systems (namely the Quantum Hall effect) was first  discussed in \cite{Son:2013rqa}.
Interactions between the torsional Newton-Cartan gravity
and matter were discussed in \cite{Jensen:2014aia}.
The correspondence between the Lifshitz space-time and
boundary torsional Newton-Cartan theory was first found
in \cite{Christensen:2013lma, Christensen:2013rfa}
for a specific Lifshitz geometry and further studied in
\cite{BHR,Hartong:2014oma, Hartong:2014pma, Hartong:2015wxa}.
In these works, the correspondence is studied by using the vielbein formalism,
in which an appropriate combination of the vielbeins and bulk gauge fields is considered.
It turns out to be very useful to use vielbeins to study the boundary theory.
This is consistent with holographic renormalization in asymptotically Lifshitz space-time,
in which the scaling dimension is calculated by using the vielbein \cite{Ross:2009ar, Ross:2011gu}.
Counter terms in Lifshitz space-time were discussed in generality in \cite{Chemissany:2014xsa}
by using the Hamilton-Jacobi formalism.

The fluid/gravity correspondence for non-relativistic fluids
has been  studied in \cite{Herzog:2008wg, Rangamani:2008gi} for a special case of
the Schr\"odinger geometry which is related to ordinary AdS by the TsT transformation.
In these studies, the non-relativistic fluids are obtained by
the light-cone reduction of relativistic fluids.
The generalization  to the charged fluid case was studied in \cite{Brattan:2010bw}.

A hydrodynamics for  Lifshitz-invariant theories was proposed in \cite{Hoyos:2013qna, Hoyos:2013cba,R}.
In this framework the velocity field is defined by a normalized Lorentz vector and the anisotropic direction of the Lifshitz symmetry depends on the frame.
The fluid appears on a surface at finite radius or on the horizon,
contrary to the Newton-Cartan theory which appears on the boundary.
The hydrodynamics proposed  contains an antisymmetric part in the hydrodynamic stress tensor that contributes a new transport coefficient to the dynamics.

In this paper, we consider the fluid/gravity correspondence for Lifshitz geometries and the relation to fluids in boundary non-relativistic theories with  Newton-Cartan symmetry.
We consider black holes in Lifshitz space-time with unbroken $U(1)$ gauge symmetry
that are solutions of the Einstein-Maxwell-dilaton (EMD) theories.
Although, the geometry has Lifshitz scaling symmetry with dynamical exponent $z$,
the bulk solution has ``charge-hyperscaling violation"\footnote{This is distinct from what is called hyperscaling violation in condensed matter physics. Our definition of charge-hyperscaling violation is based on the existence of scaling but also the existence of anomalous scaling dimensions in the charged sector. In particular, although the scaling dimensions of charge density and conductivity are canonical, the scaling of the charge density and conductivity with temperature  is controlled by the conduction exponent $\psi$, \cite{gk2,G1,gath,karch}.}   due to a nontrivial  conduction exponent $\psi$,
associated with the gauge field and the non-trivial running of the dilaton.

We consider  the black-hole solution of the theory, boost it using Galilean boosts and then we make all parameters of the solution including the velocities, $\vec x$-dependent.
We then proceed with the standard analysis introduced in  \cite{Bhattacharyya:2008jc}:
we solve the bulk equations of motion order by order in boundary derivatives
and compute and renormalize the (fluid) stress-energy tensor.
We also calculate the entropy current and consider the thermodynamic relations. What we find is as follows:

\begin{itemize}

\item The standard stress-energy tensor we obtain from the holographic calculation is expressed in terms of the fluid variables:
velocity field $v^i$, energy density $\mathcal E$ and pressure $P$,
but also contains the (particle number) density $n$ and
external source $\mathcal A_i$ associated to the $U(1)$ symmetry current.
It satisfies the condition for Lifshitz invariant theories $z\mathcal E = (d-1) P$.

\item By comparing the stress-energy tensor and the constraints from the bulk equations of motion, we find that the conservation law of the stress-energy tensor is different from that of relativistic theories but agrees with that in the Newton-Cartan theory.
In \cite{Hoyos:2013qna} the stress-energy tensor required a modification (improvement)
in order to satisfy the trace Ward identity.
Our stress-energy tensor satisfies the Ward identity without such a modification. It is Milne-boost invariant but is not gauge invariant.

\item The role of the (unbroken) U(1) symmetry in this class of theories is important. It should be noted that this U(1) symmetry is responsible for the Lifshitz background bulk solution. We find that it behaves very closely to the U(1) mass conservation symmetry in non-relativistic hydrodynamics.

\item The fluid here is non-relativistic and this is different from the
relativistic fluids analysis in \cite{Hoyos:2013qna, Hoyos:2013cba}.
Even though the continuity equation and energy conservation equation
agree with those in the ordinary non-relativistic fluids,
the Navier-Stokes equation is different from that in ordinary non-relativistic fluids.
The effects of pressure become much larger than other contributions and some terms with velocity field in the Navier-Stokes equation are absent in our result.
These absent terms are replaced by external source-dependent terms associated to the U(1) gauge field.

\item By redefining the stress-energy tensor and allowing a (Milne-invariant) Newton potential in our sources, \cite{Hartong:2014oma}-\cite{Hartong:2015wxa} we can map the fluid equations to the standard non-relativistic fluid equations coupled to the torsional Newton-Cartan geometry in the presence of a Newton potential. This is a universal result that we find interesting and far-reaching.

    Moreover, there is a stress-energy tensor that is both gauge invariant and Milne-boost invariant, but in this stress tensor the momentum density vanishes. There is also an alternative gauge invariant but Milne-boost non-invariant stress-energy tensor which agrees fully with the standard non-relativistic stress-energy tensor.

\item Our fluid can be interpreted as a non-relativistic limit of a fluid which realizes however the Lifshitz scaling symmetry.%
\footnote{The non-relativistic limit of the Lifshitz fluid
is also studied in \cite{Chapman:2014hja, Hoyos:2015lra}.}
In the ordinary non-relativistic limit of fluids, the relativistic energy is separated into
that from mass and the non-relativistic internal energy.
The non-relativistic internal energy is much smaller than the mass energy, and hence
than the relativistic energy.
In ordinary non-relativistic fluids the pressure is at the same order to the non-relativistic internal energy and hence is much smaller than the relativistic energy density.
However, in our case, pressure and the relativistic energy density are at the same order
due to the Lifshitz scaling symmetry, and
the energy density is not separated into the mass and the others.
For non-relativistic fluids with Schr\"odinger symmetry the fluid equations are  obtained by
introducing the light-cone dimensional reduction in \cite{Herzog:2008wg, Rangamani:2008gi, Brattan:2010bw}.
Instead, our non-relativistic limit arises naturally as a  rather ordinary limit.

\item We find  that the form of the fluid equations is independent of the Lifshitz exponent $z$ as well as of the (non-trivial) conduction exponent, $\psi$.
It is only the constitutive relations (equation of state) that depend on these scaling exponents.

\item The entropy satisfies the local thermodynamic relation with the energy density and pressure.
The divergence of the entropy current is non-negative, compatible with  the second law.

\end{itemize}

This paper is organized as follows.
In Section~\ref{sec:model}, we introduce the model and its solution of Lifshitz space-time.
Then, we first focus on the case with scaling exponent $z=2$.
In Section~\ref{sec:ansatz}, we introduce the hydrodynamic ansatz.
In Section~\ref{sec:solution}, we solve the equations of motion by using the derivative expansion
and obtain the solution to first order.
In Section~\ref{sec:stress}, we calculate the stress-energy tensor on the boundary
and study its  symmetries and the conservation laws.
In Section ~\ref{NC} we introduce the Newton Cartan geometry and realize it in for the boundary fluid in question.
In Section~\ref{sec:entropy}, we investigate the entropy and the thermodynamic relations.
In Section \ref{sec:GeneralA} we generalize the gauge field source in order to eventually end up with the general non-relativistic fluid equations in the presence of an external Newtonian potential.
In Section~\ref{sec:general}, we consider the generalization to general $z$.
Section~\ref{sec:conclusion} is devoted to our conclusions and further discussion.

Appendix \ref{app:notation} contains a list of the variables used in this paper and their definition. It contains also comparison of variables with two other relevant papers in the literature.
More details on the calculation for first order solution and
boundary stress-energy tensor are described in
Appendix~\ref{app:solution} and \ref{app:stress}, respectively.
In Appendix~\ref{app:generalcounter}, we consider the analysis of the  general counter terms.
In Appendix~\ref{app:RegA}, we discuss the regularity of the gauge field at the horizon.
More details on the solution and counter terms for general $z$ are discussed in
Appendices ~\ref{app:general} and \ref{app:counter}, respectively.

\section{U(1)-invariant, charge-hyperscaling violating Lifshitz theory}\label{sec:model}

We consider a holographic theory with Lifshitz scaling and an unbroken U(1) global symmetry in $d$ space-time dimensions. The dual $(d+1)$-dimensional gravity theory will have a
massless  $U(1)$ gauge field $A_\mu$ and a dilaton $\phi$.
The bulk action is given by
\begin{equation}
 S
 = \frac{1}{16\pi G} \int d^{d+1} x \sqrt{-g}
 \left(R - 2 \Lambda - \frac{1}{2} (\partial\phi)^2 - \frac{1}{4} e^{\lambda\phi}F^2\right) \ ,
\label{BulkAction}
\end{equation}
where $F = dA$ is the field strength of the gauge field and $\lambda$ is a dimensionless  coupling constant of the bulk theory.
The equations of motion are given by
\begin{align}
 R_{\mu\nu}
 &=
 \frac{2 \Lambda}{d-1} g_{\mu\nu}
 + \frac{1}{2} (\partial_\mu\phi)(\partial_\nu\phi)
 + \frac{1}{4} e^{\lambda\phi}
 \left(
  2 F_{\mu\rho} {F_\nu}^\rho - \frac{1}{d-1} e^{\lambda\phi} F^2 g_{\mu\nu}
 \right)
\label{EinsteinEq}
\\
 0
 &=
 \nabla_\mu (e^{\lambda\phi} F^{\mu\nu}) ,
\label{MaxwellEq}
\\
 \Box \phi
 &=
 \frac{1}{4} \lambda e^{\lambda\phi} F^2 \ .
\label{DilatonEOM}
\end{align}
This model has the Lifshitz geometry as a solution;
\begin{equation}
 ds^2 = - r^{2z} dt^2 + \frac{dr^2}{r^2} + \sum_i r^2 (dx^i)^2 ,
\label{1}\end{equation}
with the following gauge field and dilaton;
\begin{align}
 A_t &= a r^{z+d-1} \ , &
 e^{\lambda\phi} &= \mu r^{2(1-d)} \ .
\label{PureA}
\end{align}
The boundary is at $r\to\infty$.
The parameters $z$, $a$ and $\mu$ are related to the parameters of the action (coupling constants) as
\begin{align}
 \lambda^2 &= 2 \frac{d-1}{z-1} \ ,\label{10} \\
 \Lambda &= - \frac{(z-d-1)(z+d-2)}{2} \ , \label{11}\\
 \mu a^2 &= \frac{2(z-1)}{z+d-1} \ . \label{Constmua}
\end{align}
This solution, although well known in the context of cosmology since a long time, was first studied in holography in \cite{taylor,kachru} and was generalized in \cite{cgkkm}.

The  metric (\ref{1}) has the Lifshitz scaling symmetry
\begin{align}
 t &\to c^z t \ , &
 x^i &\to c x^i \ , &
 r &\to c^{-1} r \ ,
\label{12}\end{align}
and no hyperscaling violation ($\theta=0$).
 However, due to the running of the dilaton, the scaling of the AC conductivity is  anomalous and its scaling with temperature or frequency is controlled by the conduction exponent $\psi$, \cite{gk2,G1,gath,karch}. It is defined from the solution for $A_t$\footnote{Note that here we use a radial coordinate that is inverse to the one used in \cite{gk2}.}
\be
A_t\sim r^{z-\psi} \ .
\label{13}\ee
We will call this {\em charge-hyperscaling violation}.

The solution (\ref{PureA}) is  a solution with charge-hyperscaling violation coming from the conduction exponent, \footnote{Although the conduction exponent is usually referred to as $\zeta$, here
we refer to it as $\psi$ in order to avoid confusion with the bulk viscosity $\zeta$.}
\be
\psi = -(d-1)
\label{14}\ee
 although the hyperscaling violation exponent $\theta$ coming from the metric vanishes.

This model also has a black hole geometry as a solution \cite{taylor,cgkkm};
\begin{equation}
 ds^2 = - r^{2z} f(r) dt^2 + \frac{dr^2}{f(r) r^2} + \sum_i r^2 (dx^i)^2 ,
\label{15}\end{equation}
where
\begin{equation}
 f = 1 - \frac{r_0^{z+d-1}}{r^{z+d-1}} \ .
\label{WarpF}
\end{equation}
The Hawking temperature of the black hole is given by
\begin{equation}
 T = \frac{z+3}{4\pi} r_0^z \ .
\label{HawkingT}
\end{equation}
The gauge field and dilaton take almost the same form as in the  the zero temperature solution
\begin{align}
 A_t &= a (r^{z+d-1} - r_0^{z+d-1}) , &
 e^{\lambda\phi} &= \mu r^{2(1-d)} .
\label{16}\end{align}
but $A_t$  vanishes at the horizon for regularity.

For a general solution, the finite part of the conductivity scales as, \cite{G1}
\be
\sigma\sim  r_0^{d-3}~e^{\lambda\phi(r_0)}\sim r_0^{\psi-2}\sim T^{\psi-2\over z}
\label{17}\ee and this is controlled by the conduction exponent, $\psi$.
We observe that the temperature dependence although scaling, does not respect the natural dimension of conductivity. This justifies the name {\em charge-hyperscaling violation} for the exponent $\psi$.

In the particular case studied here, in view of (\ref{14}) and (\ref{17})  we obtain
\be
\sigma\sim T^{-{d+1\over z}} \ .
\label{18}\ee

In this paper, we focus on the case of $d=4$. Extensions to other dimensions are expected to be straightforward.

\section{Hydrodynamic ansatz}\label{sec:ansatz}

In this section, we introduce an ansatz for the geometry
which describes the physics of fluids in the boundary quantum field theory.
We use the method proposed in \cite{Bhattacharyya:2008jc}.
We will also fix the Lifshitz exponent to be $z=2$. In a later section, we will discuss other values of $z$.

In order for the regularity at the horizon to become evident,
we change to  Eddington-Finkelstein coordinates;
\begin{equation}
 ds^2 = - r^4 f dt_+^2 + 2 r dt_+ dr +  r^2 (dx^i)^2 \; ,
\end{equation}
where the null coordinate $t_+$ is defined by
\begin{equation}
 dt_+ = dt + \frac{dr}{r^3 f} \ .
\end{equation}
The gauge field becomes
\begin{equation}
 A = a \left( r^5-r_0^5 \right) dt_+ - a r^2 dr \ ,
\end{equation}
where we have fixed the $A_r=0$ gauge in the original Fefferman-Graham coordinates.
Hereafter, we always use the Eddington-Finkelstein coordinates and
$t$ will stand for the null coordinate $t_+$.

To implement the hydrodynamic ansatz, we first boost the black hole geometry.
In the case of the ordinary Schwarzschild-AdS$_5$,
the boundary field theory is a relativistic conformal field theory,
and hence the Lorentz boost is employed. This boost leaves the sources invariant.
The Lifshitz geometry, however, corresponds to the (torsional)-Newton-Cartan theory, \cite{Christensen:2013lma}-\cite{Hartong:2015wxa}.
Therefore, we perform a Galilean boost on the black hole geometry. It is this boost that now keeps the metric sources invariant.
The metric becomes
\begin{equation}
 ds^2 = - (r^4 f - v^2 r^2) dt^2 + 2 r dt dr - 2 r^2 v^i dt\, dx^i + r^2 (dx^i)^2 \ ,
\label{BoostedBH}
\end{equation}
The gauge field and dilaton are not affected by the Galilean boost;

\begin{align}
 A &= a \left( r^5-r_0^5 \right) dt - a r^2 dr \ , &
 e^{\lambda\phi} &= \mu r^{-6} .
\label{B1}\end{align}
For a homogeneous boost, $v^i=$constant,
(\ref{BoostedBH}) and (\ref{B1}) provide  an exact solution of the equations of motion.

We now replace the parameters $r_0$ and $v^i$
by slowly-varying functions $r_0(x)$ and $v^i(x)$ of the boundary coordinates $x^\mu$.
Moreover, we promote $a$, $\mu$ and
the constant part of $A_i$ ( which is usually gauged away) to space-time dependent functions.
\begin{align}
 ds^2 &= - (r^4 f - v^2(x) r^2) dt^2 + 2 r dt dr - 2 r^2 v^i(x) dt\, dx^i + r^2 (dx^i)^2
\label{BGmetric}
\\
 f &= 1 - \frac{r_0^5(x)}{r^5}
\\
 A &= a(x) \left(r^{5} - r_0^5(x) \right) dt - a(x) r^2 dr + \mathcal A_i(x) (dx^i - v^i(x) dt),
\label{BGgauge}
\\
 e^{\lambda\phi} &= \mu (x) r^{-6} .
\label{BGscalar}
\end{align}
where $\mathcal A_i(x)$ originates from the constant part of $A_i$ but
now is replaced by functions of $x^\mu$.
This is no longer a solution of the equations of motion,
and we must introduce additional correction terms;
\begin{align}
 g_{\mu\nu} &= \bar g_{\mu\nu} + h_{\mu\nu} \ , \label{Defh}\\
 A_\mu &= \bar A_\mu + a_\mu \ , \label{DefCorA}\\
 \phi &= \bar \phi + \varphi \ , \label{DefCorPhi}
\end{align}
where the background fields $\bar g_{\mu\nu}$, $\bar A_\mu$ and $\bar\phi$ are
given by \eqref{BGmetric}-\eqref{BGscalar}.

\section{The first order solution}\label{sec:solution}

In order to obtain the first order solution for the hydrodynamic ansatz,
we consider the derivative expansion.
Then, the equations of motion can be treated as ordinary differential equations with respect to $r$,
and the correction terms, $h_{\mu\nu}$, $a_\mu$ and $\varphi$, can be calculated order by order in the boundary derivative expansion.

The differential equation can be solved at any given point.
We can take this point to be $x^\mu=0$ without loss of generality.
The parameters which are replaced by slowly varying functions can be expanded  around $x^{\mu}=0$ as
\begin{align}
 v^i(x) &= v^i(0) + x^\mu (\partial_\mu v^i)(0) + \cdots , \label{vExpand}
\\
 r_0(x) &= r_0(0) + x^\mu (\partial_\mu r_0)(0) + \cdots , \label{r0Expand}
\\
 a(x) &= a(0) + x^\mu (\partial_\mu a)(0) + \cdots , \label{aExpand}
\\
 \mu(x) &= \mu(0) + x^\mu (\partial_\mu \mu)(0) + \cdots . \label{muExpand}
\end{align}
The derivative expansion of the equations of motion
gives linear differential equations
for the correction terms $h_{\mu\nu}$, $a_\mu$ and $\varphi$ to first order.
The next-to-leading terms in \eqref{vExpand}-\eqref{muExpand} are  first order and provide  the source terms in these differential equations.

Solving the (inhomogeneous) linear differential equations for the correction terms
$h_{\mu\nu}$, $a_\mu$ and $\varphi$, we obtain the first order solution in the derivative expansion.
(See Appendix~\ref{app:solution} for more details.)
The integration constants generically modify the source terms near the boundary.
These contributions are eliminated
by setting the integration constants that modify the sources to zero.
The first order solution for the metric is given by
\begin{align}
 ds^2 &= - r^4 f dt^2 + 2 r dt dr + r^2 (dx^i - v^i dt)^2
\notag\\&\quad
 + \frac{2}{3}r^2 \partial_i v^i dt^2 + r^2 F(r) \sigma_{ij} (dx^i-v^i dt) (dx^j-v^j dt) \ ,
\label{SolGeom}
\end{align}
where $\sigma_{ij}$ is the shear tensor
\begin{equation}
 \sigma_{ij} = \left(\partial_i v^j + \partial_j v^i\right)
 - \frac{2}{3} \partial_k v^k \delta_{ij} \ ,
\end{equation}
and the function $F(r)$ is given by
\begin{equation}
 F(r) = \int dr \frac{r^3-r_0^3}{r(r^5-r_0^5)}  \ .
\end{equation}
The integration constant is chosen such that $F(r)\to 0$ in $r\to\infty$.

The first order solution for the gauge field is
\begin{equation}
 A = a(x) \left[\left(r^{5} - r_0^5(x)\right)
 - \frac{1}{3}r^3 \partial_i v^i(x)\right]dt
 - a(x) r^2 dr + \mathcal A_i(x) (dx^i - v^i(x) dt) \ ,
\label{SolA}
\end{equation}
and the dilaton has no correction term, $\varphi=0$.
The equations of motion imply that \eqref{Constmua} and its derivatives
must be satisfied even after $\mu$ and $a$ are replaced by functions of $x^{\m}$.

The solution (\ref{SolGeom}) and (\ref{SolA}) solve the bulk equations of motion if the following constraints are satisfied;
\begin{align}
 0 &= \partial_t a + v^i \partial_i a - a \partial_i v^i  ,
\label{ConstA}
\\
 0 &= \partial_t r_0 + v^i \partial_i r_0 + \frac13 r_0 \partial_i v^i  ,
\label{ConstT}
\\
 0 &= \partial_t \mathcal A_i + v^j \partial_j \mathcal A_i
 + \mathcal A_j \partial_i v^j + 5 a r_0^4 \partial_i r_0 \ .
\label{ConstS}
\end{align}
These constraints are at the origin of the hydrodynamic equations we are going to derive later on.

\section{Calculation and renormalization of the boundary stress tensor}\label{sec:stress}

In this section we consider the stress-energy tensor on the boundary.
In order to study the asymptotic behavior,
we will use the vielbein formalism as this is well adapted to the Newton-Cartan geometry.

The leading order term of the induced metric near the boundary is expressed as
\begin{equation}
 \gamma_{\mu\nu} = - r^{2z} f \tau_\mu \tau_\nu + r^2 \delta_{ab} \hat e^a_\mu \hat e^b_\nu
\label{Indm}
\end{equation}
and
\begin{equation}
 \gamma^{\mu\nu} = - r^{-2z} f^{-1} \hat v^\mu \hat v^\nu + r^{-2} \delta^{ab} \hat e_a^\mu \hat e_b^\nu \ .
\label{InvIndm}
\end{equation}
where $\gamma_{\mu\nu} = g_{\mu\nu} - n_\mu n_\nu$. $g_{\mu\nu}$ is given by \eqref{BGmetric} and $n_\mu$ is the normal vector to the $dr=0$ surface.
On this surface, the vielbeins are given by
\begin{align}
 \tau_\mu dx^\mu &= dt \ ,
&
 \hat e^a_\mu dx^\mu &= dx^a - v^a dt \ ,
\label{vl}
\\
 \hat v^\mu \nabla_\mu &= \nabla_t + v^i \nabla_i \ ,
&
 \hat e_a^\mu \nabla_\mu &= \nabla_a \ .
\label{vh}
\end{align}
$\tau_{\mu}$, $\hat v^{\mu}$ and $h^{\m\n}=\hat e_a^\mu \hat e_a^\nu$ will become the basic  geometric data of Newton-Cartan geometry that is discussed in the next section.

The formulae above specify the vielbeins in a very specific frame, $\tau_{\mu}=(1, 0)$, $\hat v^{\mu}=(1, v^i)$ that depends on the velocity $v^i$ we introduced in the solution of the previous section.

We also express the gauge field in this frame as
\begin{align}
 \hat A_0 &= \hat v^\mu A_\mu \ , &
 \hat A_a &= \hat e_a^\mu A_\mu \ .
\label{Ahat}
\end{align}
As will see in the next section, the holographic data above will correspond to an infinite set of Newton-Cartan data, related by Milne boosts.
Because of this, the holographic data above will be ``Milne boost invariant".

In order to renormalize the expectation values,
we have to add  counter terms to the action;
\begin{equation}
 S_r = S + S_\text{ct} \ .
\end{equation}
with
\begin{equation}
 S_\text{ct} =
 \frac{1}{16 \pi G }\int d^4 x \sqrt{-\gamma}
 \left(
 - 7 + \frac{5}{2} e^{\lambda\phi}\gamma^{\mu\nu} A_\mu A_\nu
 \right) \ .
\end{equation}
The variation of the renormalized action is expressed as
\begin{equation}
 \delta S_r
 = \int \left(
 - \hat S^0_\mu \delta \hat v^\mu + \hat S^a_\mu \delta \hat e_a^\mu
 + \hat J^0 \delta\hat A_0 + \hat J^a \delta\hat A_a
 + \mathcal O_\phi \delta \phi
 \right) \ .
\label{VariationOfAction}
\end{equation}
Then, the renormalized boundary theory stress-energy tensor and current are given by
\begin{align}
 \widehat T^\mu{}_\nu &= \lim_{r\to \infty} r^5 T_r^\mu{}_\nu \ ,
\label{T_hat}
\\
 J^\mu &= \lim_{r\to \infty} r^5 J_r^\mu \ .
\label{Jdef}
\end{align}
where
\begin{align}
 T_r^\mu{}_\nu &= \hat S^0_\nu \hat v^\mu - \hat S^a_\nu \hat e_a^\mu \ ,
\\
 J_r^\mu &= \hat J^0 \hat v^\mu + \hat J^a \hat e_a^\mu \ .
\end{align}
It should be noted that the stress-energy tensor $\widehat T^\mu{}_\nu$ is not a symmetric tensor. It is also {\em not gauge invariant} because we work in the vielbein formalism. As we will see later on, when we introduce Milne-boosts, it will be Milne-boost invariant. In section \ref{sec:GeneralA} we will define different stress-energy tensors with different properties under gauge transformations and Milne boosts.

The stress-energy tensor (\ref{T_hat}) is related to the ordinary Brown-York tensor as
\begin{align}
 T_r^\mu{}_\nu &= T_\text{(BY)}{}^\mu{}_\nu + J^\mu A_\nu +T_\text{(ct)}{}^\mu{}_\nu\ ,
\label{8}\end{align}
where $T_\text{(ct)}{}^\mu{}_\nu$ is the counter term contribution and
$T^{\mu\nu}_\text{(BY)}$ is the Brown-York tensor,
which can be expressed in terms of the extrinsic curvature $K_{\mu\nu}$ as
\begin{equation}
 T^{\mu\nu}_\text{(BY)} = \frac{1}{8\pi G}\left(\gamma^{\mu\nu} K - K^{\mu\nu}\right) \ .
\label{BYtensor}
\end{equation}
$J^\mu$ is calculated as
\begin{equation}
 J^\mu = \frac{1}{\sqrt{-\gamma}} \frac{\delta S}{\delta A_\mu}
 = e^{\lambda \phi} n_\nu F^{\mu\nu}  \ ,
\label{Current}
\end{equation}
where $n^\mu$ is the unit normal to the boundary.

The renormalized stress-energy tensor is obtained from (\ref{T_hat}), (\ref{Jdef}) as
\begin{align}
 \widehat T^0{}_0
 &=
 \frac{1}{8\pi G}  \left(-\frac{3}{2} r_0^5 - \frac{1}{a} v^i \mathcal A_i \right)
\ ,
\label{T00}
\\
 \widehat T^i{}_0
 &=
 \frac{1}{8\pi G}  \left(- \frac{5}{2} r_0^5 v^i + \frac{1}{2} \partial_i r_0^5 - \frac{1}{a} v^i v^j \mathcal A_j
 + \frac{1}{2} r_0^3 \sigma_{ij} v^j  \right)
  \ ,
\label{Ti0}
\\
  \widehat T^0{}_i
 &=
 \frac{1}{8\pi G} \frac{1}{a} \mathcal A_i  \ ,
\label{T0i}
\\
  \widehat T^i{}_j
 &=
 \frac{1}{8\pi G}  \left( r_0^5 \delta_{ij} - \frac{1}{2} r_0^3 \sigma_{ij}
 + \frac{1}{a} v^i \mathcal A_j \right)
 \ .
\label{Tij}
\end{align}
The above expressions show that the stress-energy tensor $\widehat T^\mu{}_\nu$
contains the gauge field $\mathcal A_\mu$ and hence is not gauge invariant.
We will discuss several other definitions of gauge invariant stress-energy tensors
in section~\ref{sec:GeneralA}.

\subsection{Energy and momentum conservation}

We now consider the conservation of the stress-energy tensor.

In the Newton-Cartan theory (that is described in more detail in the next section), the conservation law takes a slightly different form from standard relativistic cases.
It cannot be expressed in a unified form in terms of the space-time stress-energy tensor
and we have to introduce the energy vector $\widehat{\mathcal E}^\mu$,
momentum density $\widehat{\mathcal P}_\mu$ and stress tensor $\widehat{\mathcal T}^\mu{}_\nu$,
which are defined by
\begin{align}
 \mathcal {\widehat E}^\mu
 &= - \widehat T^\mu{}_\nu \hat v^\nu \ ,
\label{Edef}
\\
 \mathcal {\widehat P}_\mu
 &=
 \widehat T^\rho{}_\nu \tau_\rho \hat e_a{}^\nu \hat e^a{}_\mu  \ ,
\label{Pdef}
\\
 \mathcal {\widehat T}^\mu{}_\nu
 &=
 \widehat T^\rho{}_\sigma (\hat e^a{}_\rho \, \hat e_a{}^\mu) (\hat e_b{}^\sigma \, \hat e^b{}_\nu) \ .
\label{Tdef}
\end{align}
Then, the conservation of energy and momentum is given by (see for example \cite{Jensen:2014aia,Hartong:2014oma,Hartong:2015wxa})
\begin{align}
 \nabla_\mu \widehat{\mathcal E}^\mu &=
 - \frac{1}{2} (\nabla^\mu \hat v^\nu + \nabla^\nu \hat v^\mu) \widehat{\mathcal T}_{\mu\nu} \ ,
\label{ConsT}
\\
 \nabla_\mu \widehat{\mathcal T}^\mu{}_i &= \hat v^\mu \nabla_i \widehat{\mathcal P}_\mu
 - \nabla_\mu (\hat v^\mu \widehat{\mathcal P}_i) \ .
\label{ConsS}
\end{align}
{}From the first order solution, \eqref{T00}-\eqref{Tij},
the energy vector $\widehat{\mathcal E}^\mu$,
momentum vector (1-form) $\widehat{\mathcal P}_\mu$ and stress tensor $\widehat{\mathcal T}^\mu{}_\nu$
are given by
\begin{align}
 \widehat{\mathcal E}^0
 &=
 \frac{3}{16\pi G} r_0^5 \ ,
&
 \widehat{\mathcal E}^i
 &=
 \frac{1}{16\pi G}\left(3 r_0^5 v^i - \partial_i r_0^5 \right)
\ ,
\label{EnergyCurrent}
\\
 \widehat{\mathcal P}_0
 &=
 - \frac{1}{16\pi G a} v^i \mathcal A_i \ ,
&
 \widehat{\mathcal P}_i
 &=
 \frac{1}{16\pi G a} \mathcal A_i \ ,
\\
 \widehat{\mathcal T}^i{}_j
 &= \frac{1}{8\pi G} r_0^5 \delta_{ij} - \frac{1}{16 \pi G} r_0^3 \sigma_{ij} \ ,
&
 \widehat{\mathcal T}^i{}_0
 &=
 - \frac{1}{8\pi G} r_0^5 v^i + \frac{1}{16\pi G} r_0^3 v^j \sigma_{ij}
\ . \label{StressTensor}
\end{align}
The other components of $\widehat {\mathcal T}^\mu{}_\nu$, namely $\widehat {\mathcal T}^0{}_0$ and $\widehat {\mathcal T}^0{}_i$  vanish.
The Lifshitz-scaling invariance condition becomes
\begin{equation}
 z \tau_\mu \hat v^\nu \widehat T^\mu{}_\nu + \hat e^a_\mu \hat e_a^\nu \widehat T^\mu{}_\nu = 0 \ .
\end{equation}
in terms of the stress-energy tensor $\widehat T^\mu{}_\nu$,
or equivalently
\be
z\widehat{\mathcal E}^0- \widehat{\mathcal T}^i{}_i=0
\ee

As was already noted, the momentum density contains
a contribution from the external source $\mathcal A_\mu$
and hence is not gauge invariant.
The external source $\mathcal A_\mu$ dependence originates from
the Newton-Cartan definition of the stress-energy tensor.
Introducing an appropriate redefinition of the stress-energy tensor,
we will obtain a gauge-invariant momentum density,
which is related to the velocity field $v^i$.
We will discuss this in section~\ref{sec:GeneralA}.

Using \eqref{EnergyCurrent}-\eqref{StressTensor},
the energy conservation \eqref{ConsT} becomes
\begin{equation}
  0 = \frac{1}{2}\left[15 r_0^4 \partial_t r_0 + 15 r_0^4 v^i \partial_i r_0 + 5 r_0^5 \partial_i v^i
 - 5 r_0^4 \partial^2 r_0 - 20 (\partial_i r_0)^2 - \frac{1}{2} r_0^3 (\sigma_{ij})^2 \right]\ ,
\label{SolConsT}
\end{equation}
and its leading order terms give
\begin{align}
 \frac{5}{2} \left(
  3 r_0^4 \partial_t r_0 + 3 r_0^4 v^i \partial_i r_0 + r_0^5 \partial_i v^i
 \right)
 &= 0 ,
\end{align}
which is the same as the constraint \eqref{ConstT}.
The momentum conservation \eqref{ConsS} becomes
\begin{align}
 0 &=
 r_0^4 \partial_i r_0 + \frac{1}{a} \partial_t \mathcal A_i + \frac{1}{a} v^j \partial_j \mathcal A_i
 + \frac{1}{a} \mathcal A_j \partial_i v^j
 - \frac{1}{2} \partial_j (r_0^3 \sigma_{ij})
 + \mathcal A_i \left[\partial_t \left(\frac{1}{a}\right) + \partial_j \left(\frac{v^j}{a}\right)\right]
\ . \label{SolConsS}
\end{align}
The leading order terms give
\begin{align}
 5 r_0^4 \partial_i r_0 + \frac{1}{a} \mathcal A_j \partial_i v^j
 + \frac{1}{a} \partial_t \mathcal A_i + \frac{1}{a} v^j \partial_j \mathcal A_i
 + \mathcal A_i \left[\partial_t \left(\frac{1}{a}\right) + \partial_j \left(\frac{v^j}{a}\right)\right]
 &= 0
\end{align}
which is a combination of \eqref{ConstA} and \eqref{ConstS}.

The next-to-leading order terms of the conservation law provide
the constraints at second order.
In order to calculate the solution to second order,
we need to introduce the derivative expansion of the correction terms,
for example,
\begin{align}
 g_{\mu\nu} = \bar g_{\mu\nu} + \epsilon h_{\mu\nu}^{(1)}
 + \epsilon^2 h_{\mu\nu}^{(2)} + \cdots \ .
\end{align}
where $\bar g_{\mu\nu}$ is given by \eqref{BGmetric} and
$h_{\mu\nu}^{(1)}$ is the correction terms which we calculated in the previous section.
The expansion parameter $\epsilon$ is that of the derivative expansion,
$\partial_\mu = \mathcal O(\epsilon)$.
In order to calculate the second order solution, we further introduce
the second order correction terms $h_{\mu\nu}^{(2)}$.
However, as the correction terms do not contribute to the constraint at  first order,
these second order correction terms do not contribute to the constraint at the second order.
Therefore, we do not need to take $h_{\mu\nu}^{(2)}$ into account
to study the second order constraints.

The background metric $\bar g_{\mu\nu}$ does not consist only of $\mathcal O(\epsilon^0)$ contributions,
but also contains higher order corrections.
The higher order corrections in $\bar g_{\mu\nu}$ are included
in $x$-dependent parameters $r_0$, $v^i$, $a$, and $\mathcal A_i$.
They can be expanded as
\begin{align}
 r_0(x)
 &= r_0^{(0)}(x) + \epsilon r_0^{(1)}(x) + \cdots \ ,
\\
 v^i(x)
 &= v^{i\,(0)}(x) + \epsilon v^{i\,(1)}(x) + \cdots \ ,
\\
 a(x)
 &= a^{(0)}(x) + \epsilon a^{(1)}(x) + \cdots \ ,
\\
 \mathcal A_i(x)
 &= \mathcal A_i^{(0)}(x) + \epsilon \mathcal A_i^{(1)}(x) + \cdots \ .
\end{align}
where $r_0^{(0)}$, etc.\ are the leading order terms which we studied in the previous section,
and satisfies the constraints at the first order.
The higher order terms, for example $r_0^{(1)}$, do not contribute the first order terms,
but must satisfy the second order constraints.

After some algebra, the second order constraint equation for the gauge field gives
\begin{equation}
 0 = \partial_t a^{(1)} + v^i{}^{(0)} \partial_i a^{(1)} + v^i{}^{(1)} \partial_i a^{(0)}
 - a^{(0)} \partial_i v^i{}^{(1)} - a^{(1)} \partial_i v^i{}^{(0)} \ .
\end{equation}
Together with the first order constraint,
it can be expressed as
\begin{equation}
 0 = \partial_t a + v^i \partial_i a - a \partial_i v^i
\label{Const2A}
\end{equation}
This implies that there are no additional terms in this constraint at second order. In a similar fashion, from the spatial component of the constraints in Einstein equation
we obtain
\begin{align}
 0 &=
 - \partial_i r_0^5 - \frac{1}{a} \partial_t A_i - \frac{1}{a} v^j \partial_j A_i
 - \frac{1}{a} A_j \partial_i v^j
 + \frac{1}{2} \partial_j (r_0^3 \sigma_{ij}) \ .
\label{Const2S}
\end{align}
{}From the temporal component, we obtain
\begin{align}
 0 &= 15 r_0^4 \partial_t r_0 + 15 r_0^4 v^i \partial_i r_0 + 5 r_0^5 \partial_i v^i
 - 5 r_0^4 \partial^2 r_0 - 20 (\partial_i r_0)^2 - \frac{1}{2} r_0^3 (\sigma_{ij})^2 \ .
\label{Const2T}
\end{align}
\eqref{Const2T} agrees with \eqref{SolConsT}, and
an appropriate combination of \eqref{Const2A} and \eqref{Const2S}
gives \eqref{SolConsS}.

{}From the stress-energy tensor and the current, we can read off
the energy density $\mathcal E$, charge density $n$ and pressure $P$ as
\begin{align}
 \mathcal E &= \frac{3}{16\pi G} r_0^5 \ , &
 n &= \frac{1}{16\pi G a} \ , &
 P &= \frac{1}{8\pi G} r_0^5 \ .
\label{FluidVariables}
\end{align}
In terms of these quantities,
the energy flow, momentum density and stress tensor are expressed as
\begin{align}
 \widehat{\mathcal E}^0 &= \mathcal E \ ,
\\
 \widehat{\mathcal E}^i &= \mathcal E v^i - \kappa \partial_i T \ ,
\label{KappaInt}
\\
 \widehat{\mathcal P}_i &= n \mathcal A_i \ ,
\\
 \widehat{\mathcal T}^i{}_j &= P \delta_{ij} - \eta \sigma_{ij} \ ,
\label{EtaInt}
\end{align}
where $\kappa$ is the heat conductivity and $\eta$ is the shear viscosity whose values are
\begin{align}
 \kappa &= \frac{1}{8\pi G} r_0^3 \ , &
 \eta &= \frac{1}{16\pi G} r_0^3 \ .
\end{align}
In terms of the temperature (note that here $z=2$)
\begin{equation}
 T = \frac{5}{4\pi} r_0^2 \ ,
\end{equation}
fluid variables and transport coefficients can be expressed as
\begin{align}
 \mathcal E &= \frac{3}{16\pi G} \left(\frac{4\pi}{5} T\right)^{5/2} \ , &
 P &= \frac{1}{8\pi G} \left(\frac{4\pi}{5} T\right)^{5/2} \ , &
\end{align}
and
\begin{align}
 \kappa &= \frac{1}{8\pi G} \left(\frac{4\pi}{5} T\right)^{3/2} \ , &
 \eta &= \frac{1}{16\pi G} \left(\frac{4\pi}{5} T\right)^{3/2} \ .
\end{align}

The scaling dimension under the Lifshitz scaling is given as
\begin{align}
 [\mathcal E] &= 5 \ , &
 [P] &= 5 \ , \\
 [n] &= 3 \ , &
 [T] &= 2 \ , \\
 [v^i] &= 1 \ , &
 [\mathcal A_i] &= 1 \ , \\
 [\kappa] &= 3 \ , &
 [\eta] &= 3 \ .
\end{align}
The Lifshitz invariance condition now becomes
\begin{equation}
 z \mathcal E = (d-1) P \ .
\end{equation}

In terms of the above fluid variables and transport coefficients,
the fluid equations take the following form;
\begin{align}
 0 &= \partial_t \mathcal E + v^i \partial_i \mathcal E + (\mathcal E + P) \partial_i v^i
 - \frac{1}{2} \eta \sigma_{ij}\sigma_{ij}
 - \partial_i( \kappa \partial_i T) \ , \label{LifContinuity}
\\
 0 &= \partial_i P + n \partial_t \mathcal A_i
 + n v^j \partial_j \mathcal A_i + n \mathcal A_j \partial_i v^j
  - \partial_j \left(\eta\sigma_{ij} \right)
\ , \label{LifNavierStokes}
\\
 0 &= \partial_t n + \partial_j (n v^j) \ . \label{LifChargeCons}
\end{align}
Eq.~\eqref{LifNavierStokes} can also be expressed as
\begin{equation}
 \partial_i P  - \partial_j \left(\eta\sigma_{ij} \right)
 = \mathcal F_{i\mu} J^\mu \label{LifNavierStokesF}
\end{equation}
where the current is given by
\begin{equation}
 J^\mu = n \hat v^\mu \ .
\label{new4}\end{equation}
$\hat v^{\m}$ is defined in (\ref{vh}) and the field strength is defined as $\mathcal F = d\mathcal A$ with
\begin{equation}
 \mathcal A = \mathcal A_i (dx^i - v^i dt) \ . \label{gaugeBH}
\end{equation}

For comparison,
the ordinary non-relativistic fluid equations are given by
\begin{align}
 0 &= \partial_t \mathcal E + v^i \partial_i \mathcal E + (\mathcal E + P) \partial_i v^i
 - \frac{1}{2} \eta \sigma_{ij}\sigma_{ij}
 - \partial_i( \kappa \partial_i T) \ , \label{NREnergy}
\\
 0 &= \partial_i P + n \partial_t v^i
 + n v^j \partial_j v^i
 - \partial_j \left(\eta\sigma_{ij} \right)
\ , \label{NRNavierStokes}
\\
 0 &= \partial_t n + \partial_j (n v^j) \ , \label{NRContinuity}
\end{align}
where \eqref{NREnergy} gives the conservation of energy,
the Navier-Stokes equation \eqref{NRNavierStokes} comes from conservation of momentum,
and the continuity equation \eqref{NRContinuity} implies the conservation of mass density.
Eqs.~\eqref{LifContinuity} and \eqref{LifChargeCons} agree with
the energy conservation \eqref{NREnergy} and continuity equation \eqref{NRContinuity}, respectively.
However, \eqref{LifNavierStokes} is different from the Navier-Stokes equation \eqref{NRNavierStokes}. Before we proceed further, we must clarify the role of Newton Cartan theory.

\section{Newton Cartan theory and Milne-boost invariance\label{NC}}

We  briefly review here the Newton-Cartan theory \cite{kunzle, Duval:1983pb}.
We first introduce the Galilei metric of $d$-dimensional Galilei space-time,
which consists of a 1-form $\tau_\mu$ and a contravariant symmetric tensor $h^{\mu\nu}$ of rank $(d-1)$.
The 1-form $\tau$ defines the time direction of the Galilei space-time
and $h^{\mu\nu}$ gives the spatial inverse metric.
They satisfy the orthogonality condition,
\begin{equation}
 \tau_\mu h^{\mu\nu} = 0 \ .
\end{equation}

The Galilei data $(\tau_\mu, h^{\mu\nu})$ are constant under the covariant derivative;
\begin{align}
 \nabla_\nu \tau_\mu &= 0 \ , &
 \nabla_\rho h^{\mu\nu} &= 0 \ .
\label{NCConstMetric}
\end{align}
Contrary to  Einstein gravity,
\eqref{NCConstMetric} does not uniquely fix the Galilei connection.
In order to determine the connection, we must introduce
a contravariant vector $\bar v^\mu$ (not to be confused with velocities) and a two-form $\mathcal B_{\m\n}$.
The vector $\bar v^\mu$ satisfies the normalization condition
\begin{equation}
 \tau_\mu \bar v^\mu = 1 \ .
 \label{tu}
\end{equation}
By using the vector $\bar v^\mu$, we also define
the spatial covariant (symmetric) metric $\bar h_{\mu\nu}$, which satisfies
\begin{align}
 \bar h_{\mu\nu} \bar v^\mu &= 0 \ , &
 \bar h_{\mu\rho} h^{\rho\nu} &= \bar P_\mu{}^\nu \equiv \delta_\mu{}^\nu - \tau_\mu \bar v^\nu \ .
\label{defhbar}
\end{align}
Then, the Newton-Cartan connection can be  expressed as
\begin{equation}
 \Gamma^\rho_{\mu\nu}
 = \bar v^\rho \partial_\mu \tau_\nu
 + \frac{1}{2} h^{\rho \sigma}
 \left(\partial_\mu \bar h_{\nu\sigma} + \partial_\nu \bar h_{\mu\sigma}
  - \partial_\sigma \bar h_{\mu\nu} \right)
 + \frac{1}{2} h^{\rho \sigma}
 \left(\tau_\mu {\mathcal B}_{\nu \sigma}
  + \tau_\nu {\mathcal B}_{\mu \sigma}\right) \ .
\label{NCconnection}
\end{equation}
In general, the Newton-Cartan connection \eqref{NCconnection} has  torsion,
\be
{T^{\rho}}_{\m\n}= \Gamma^\rho_{\mu\nu}- \Gamma^\rho_{\nu\mu}=\bar v^{\rho}(\pa_{\m}\tau_{\n}-\pa_{\n}\tau_{\m})
\label{tors}\ee
The curvature is defined via the commutator of the covariant derivative and given by
\begin{equation}
 \mathcal R^\mu{}_{\nu\rho\sigma}
 =
 \partial_\rho \Gamma^\mu_{\nu\sigma} -  \partial_\sigma \Gamma^\mu_{\nu\rho}
 + \Gamma^\mu_{\alpha\rho} \Gamma^\alpha_{\nu\sigma}
 - \Gamma^\mu_{\alpha\sigma} \Gamma^\alpha_{\nu\rho} \ .
\end{equation}
If we impose the Newtonian condition
\begin{equation}
 \mathcal R^{[\mu}{}_{(\nu}{}^{\rho]}{}_{\sigma)} = 0 \ ,
\label{NewtonCond}
\end{equation}
where $[\cdots]$ and $(\cdots)$ in the indices stand for
the antisymmetric part and symmetric part, respectively,
we obtain the condition $d{\mathcal B}=0$.
Then, ${\mathcal B}$ is (locally) the field strength of a gauge field;
${\mathcal B} = d{ B}$.

To summarize, the Newton-Cartan data that determine a given Newton-Cartan frame are $(h^{\m\n}, \tau_{\m},\bar v^{\m},B_{\m})$.

It should be noted that $\bar v^\mu$ has no a priori relation to the fluid velocity,
and is in general, different from the fluid velocity vector $\hat v^\mu$,
although $\bar v^\mu$ is referred to as the ``velocity field" sometimes in the  Newton-Cartan literature.
Here,  $\bar v^\mu$
is the inverse timelike vielbein  to be distinguished in general from the fluid velocity field.

Different Newton-Cartan data $(h^{\m\n}, \tau_{\m},\bar v^{\m},B_{\m})$, may describe the same physics. To see this, we introduce the concept of the Milne boost (see \cite{Duval:1993pe}), which is an internal symmetry of the Newton-Cartan theory.
Here, we focus on the torsion-free cases, since for our solution on the  gravity side, (\ref{Indm}), (\ref{InvIndm})
we have  $\tau_\mu = (1,0)$ and this give zero torsion in (\ref{tors}).
\footnote{
The Milne boost in torsional cases is discussed in \cite{Jensen:2014aia,Jensen:2014ama}.
}

We introduced $\bar v^\mu$ and ${ B}_\mu$ to define the Newton-Cartan connection. Two pairs
$(\bar v^\mu, { B}_\mu)$ and $(\bar v'{}^\mu, { B}'_\mu)$
are physically the same if they give the same Newton-Cartan connection.
The torsionless part of the Newton-Cartan connection is invariant under
the following (Milne boost) transformation;
\begin{align}
 \bar v^\mu &\to \bar v'{}^\mu = \bar v^\mu + h^{\mu\nu} V_\nu \ , \label{MilneV}\\
{B} &\to { B}' =
 { B} + \bar P_\mu^\nu V_\nu dx^\mu
 - \frac{1}{2} h^{\mu\nu} V_\mu V_\nu \tau_\rho dx^\rho \ , \label{MilneA}
\end{align}
\be
\bar h'_{\m\n}=\bar h_{\m\n}-(\tau_{\m}{\bar P_{\n}}^{\rho}+\tau_{\n}{\bar P_{\m}}^{\rho})V_{\rho}+\tau_{\m}\tau_{\n}h^{\r\s}V_{\rho}V_{\sigma}
\label{MilneV1}\ee
where $V_\nu$ is a vector  which parametrizes the Milne-boost  transformation.

It should be noted that all non-trivial degrees of freedom of $\bar v^\mu$ can be
absorbed into ${B}$ by using the Milne boost.
The normal direction to the timeslice is fixed by the normalization condition
$\tau_\mu \bar v^\mu = 1$ and the other directions are freely transformed by
the Milne boost \eqref{MilneV}.
Therefore, we may choose an arbitrary but appropriately normalized inverse timelike vielbein $\bar v^\mu$.

Now, we will relate the Newton-Cartan data to the vielbeins \eqref{vl} and \eqref{vh} which we introduced
in the induced metric on the boundary \eqref{Indm} and \eqref{InvIndm}.
The timelike vielbein $\tau_\mu$ is simply identified to
that in the Newton-Cartan theory.
The inverse spacelike vielbein $\hat e_a^\mu$ should also be identified with
that in the Newton-Cartan theory,
which implies that the inverse spatial metric in Newton-Cartan theory is
expressed in terms of $\hat e_a^\mu$ as
\begin{equation}
 h^{\mu\nu} = \hat e_a^\mu \hat e_a^\nu \ .
\end{equation}
In the frame we use, it is the unit matrix in the spatial directions.
For the leftover Newton-Cartan data $(\bar v^{\m},B_{\m})$ there are many choices related by Milne boosts.

By using the Milne boost (\ref{MilneV}),
we can choose a special ``frame'' in which the inverse timelike vielbein
equals the fluid velocity $\bar v'{}^\mu = \hat v^\mu$.%
\footnote{
Here, ``frame'' is different from the coordinate frame but
a special point in the internal space of the Milne boost symmetry.
No coordinate transformation is needed to take this ``frame.''
}
We will call this Newton-Cartan frame the ``holographic frame" from now on.
It remains to identify the gauge field $B_{\m}$ in this frame. This is facilitated by comparing our equation in (\ref{LifNavierStokesF})
with the one derived in \cite{Jensen:2014aia} in Newton Cartan theory (equation (5.17) of that paper). This gives the following identification, $B={\mathcal A}$ with
\be
{\mathcal A}=(-v^i{\mathcal A}_i,{\mathcal A}_i)\sp B_t=-v^i{\mathcal A}_i\sp B_i= {\mathcal A}_i
\label{BA}\ee
where ${\mathcal A}_i$ appears in  (\ref{T00})-(\ref{Tij}).
We conclude, that in the ``holographic frame" the Newton-Cartan data are
\be
\tau_{\m}= (1,\vec 0)\sp h^{\m\n}=\left(\begin{matrix}0&0&0&0\\0&1&0&0\\0&0&1&0\\0&0&0&1\end{matrix}\right)\sp \bar v^{\mu}\to \hat v^{\mu}=(1,\vec v)
\ee

together with (\ref{BA}).
In any other frame, $\tau^{\mu}$ and $h^{\m\n}$ remain invariant, but $\bar v^{\mu}$ and $B_{\m}$ change by the Milne boosts, (\ref{MilneV})

We would like now to change to a more canonical (flat) frame that is appropriate for non-relativistic physics, in particular the standard Navier-Stokes equation. We will call this the ``Newton frame" and it is determined by $\bar v^{\m}=(1,\vec 0)$.
We will go from the holographic frame to the Newton frame by a Milne boost with parameter $V_{\m}=(0,-\vec v)$. In the Newton frame we therefore obtain a new gauge field that we call $\widetilde{\mathcal A}$ using  (\ref{MilneV})
\be
\bar v^{\m}=(1,\vec 0)\sp \widetilde{\mathcal A}={\mathcal A}- v^i dx^i + \frac{1}{2} v^2 dt=({\mathcal A}_i-v^i)dx^i-\left(v^i {\mathcal A}_i-{1\over 2}v^2\right)dt \ .
\label{gaugeNC}
\end{equation}
It should be noted that $\hat v^\mu$ does not transform under the Milne boost.
The inverse vielbein $\bar v^\mu$ transforms under the Milne boost, but
the fluid velocity $\hat v^\mu$ is Milne-boost invariant.

We will define also a class of gauge fields that are Milne-boost invariant. It is simple to show that for any Milne-invariant vector $X^{\m}$, that is normalized: $\tau_{\m}X^{\m}=1$, the following gauge field
\be
\widehat B=B
 + \bar h_{\mu\nu} X^\nu dx^\mu
 - \frac{1}{2}\bar h_{\mu\nu} X^\mu X^\nu \tau_\rho dx^\rho \ .
\label{new1}\ee
is Milne-boost invariant  as can be directly verified using the transformations in (\ref{MilneV})-(\ref{MilneV1}).

We choose as such a vector the fluid velocity vector, $X^{\mu}=\hat v^{\mu}$, which satisfies $\tau_{\m}\hat v^{\m}=1$, to define the invariant gauge field as in (\ref{new1})
\be
B_{inv}=B
 + \bar h_{\mu\nu}\hat v^\nu dx^\mu
 - \frac{1}{2}\bar h_{\mu\nu} \hat v^\mu \hat v^\nu \tau_\rho dx^\rho \ .
\label{new12}\ee
We can evaluate $B_{inv}$ in the Newton frame (\ref{gaugeNC}) to find that
\be
B_{inv}=\widetilde {\mathcal A}+ \bar h_{\mu\nu}\hat v^\nu dx^\mu
 - \frac{1}{2}\bar h_{\mu\nu} \hat v^\mu \hat v^\nu \tau_\rho dx^\rho=\widetilde {\mathcal A}+ v^i dx^i - \frac{1}{2} v^2 dt={\mathcal A}
\ee
As $B_{inv}$ is Milne-boost invariant, its evaluation in the holographic frame will also give the same result. We conclude that the gauge field $\mathcal A$ in the  holographic frame, given in (\ref{BA}) is Milne-boost invariant.

Next, we consider the Navier-Stokes equation.
We have already found  the Navier-Stokes equation
in the holographic frame in (\ref{LifNavierStokes}) to have the form
\begin{equation}
{\mathcal F}_{i\mu} J^\mu = \partial_i P - \partial_j \left(\eta\sigma_{ij} \right)
\ , \label{NCNavierStokes}
\end{equation}
where ${\mathcal F} = d{\mathcal A}$ and where as we have shown above, all quantities that enter are Milne-boost invariant.
However this does not look like the usual non-relativistic Navier-Stokes equation (\ref{NRNavierStokes}) because we are not in the Newton frame.
To do this we must rewrite it using the gauge field $\widetilde{\mathcal A}$ in the Newton frame, (\ref{gaugeNC}).
We directly compute
\be
\widetilde{\mathcal F}_{i\m}J^{\mu}=\mathcal F_{i\m}J^{\m}+n\pa_t v^i+nv^j\pa_i v^j
\ee
using (\ref{LifNavierStokes}), (\ref{new4}) and (\ref{gaugeNC}).
Substituting this equation in (\ref{NCNavierStokes}) we obtain
\begin{equation}
 \partial_i P + n \partial_t v^i
 + n v^j \partial_j v^i
 - \partial_j \left(\eta\sigma_{ij} \right) = \widetilde{\mathcal F}_{i\mu} J^\mu \ .
\label{new5}\end{equation}
which is the conventional Navier Stokes equation albeit in the presence of an external force.

To bring this equation to an even more familiar form we follow \cite{Hartong:2015wxa}, and choose the gauge field $\mathcal A$ such that
\begin{equation}
 \hat v^\mu = \delta^{\mu}_t + h^{\mu\nu}\mathcal A_\nu ~~~\to~~~{\mathcal A}_i=v^i\ . \label{ansatzA}
\end{equation}
This implies
\begin{align}
 \widetilde{\mathcal A}_t &= - \frac{1}{2} v^2 \ , &
 \widetilde{\mathcal A}_i &= 0 \ .
 \label{RestPotential}
\end{align}

Then, \eqref{new5} is expressed as

\begin{equation}
 \partial_i P + n \partial_t v^i
 + n v^j \partial_j v^i
 - \partial_j \left(\eta\sigma_{ij} \right) = n \partial_i \widetilde\Phi \ .
\label{NavierStokesNewton}
\end{equation}
This expression agrees with the Navier-Stokes equation with the external force $F_i$;
\begin{equation}
 \partial_i P + n \partial_t v^i
 + n v^j \partial_j v^i
 - \partial_j \left(\eta\sigma_{ij} \right) = F_i \ ,
\label{NavierStokesForce}
\end{equation}
the force being the gravitational force from the Newton potential $F_i = n\partial_i \widetilde\Phi$.
Therefore, \eqref{new5} can be interpreted as
the Navier-Stokes equation in the non-trivial Newton potential;
\begin{equation}
 \widetilde\Phi =  \widetilde{\mathcal A}_t = - \frac{1}{2} v^2 \ .
\label{NewtonPotential}
\end{equation}

The energy conservation \eqref{NREnergy} is for the ``internal energy" and therefore is  not affected by the Newton potential.
The conservation of total energy is obtained by appropriate combination
of \eqref{LifContinuity} and \eqref{new5},
and can be expressed in terms of the Newton potential \eqref{NewtonPotential} as
\begin{equation}
 0 = \partial_t \left(\mathcal E + \frac{1}{2} n v^2 - n\widetilde\Phi\right)
 + \partial_i \left[\left(\mathcal E + P + \frac{1}{2} n v^2 - n\widetilde\Phi\right)v^i
  - \eta \sigma_{ij} v^j - \kappa \partial_i T \right] + n \partial_t\widetilde\Phi\ .
\label{EnergyConsNewton}
\end{equation}
This is consistent with the energy conservation of fluids in the Newton potential;
for time-independent Newton potential $\partial_t \widetilde\Phi = 0 $,
the energy conservation is expressed as
\begin{equation}
 0 = \partial_t \left(\mathcal E + \frac{1}{2} n v^2 - n\widetilde\Phi\right)
 + \partial_i \left[\left(\mathcal E + P + \frac{1}{2} n v^2 - n\widetilde\Phi\right)v^i
  - \eta \sigma_{ij} v^j - \kappa \partial_i T \right] \ .
\end{equation}

The stress-energy tensor \eqref{T00}-\eqref{Tij} is now rewritten as
\begin{align}
  \widehat T^0{}_0
 &=
 - \left(\mathcal E - n \widetilde\Phi + \frac{1}{2} n v^2 \right) \ ,
\\
  \widehat T^i{}_0
 &=
 - \left(\mathcal E + P - n \widetilde\Phi + \frac{1}{2} n v^2 \right) v^i
 + \eta\sigma_{ij} v^j + \kappa \partial_i T \ ,
\\
 \widehat T^0{}_i
 &=
 n v^i \ ,
\\
 \widehat T^i{}_j
 &=
 P \delta_{ij} - \eta \sigma_{ij}
 + n v^i v^j \ .
\end{align}
This is the ordinary stress-energy tensor for non-relativistic fluids
with the Newton potential terms.
The conservation of total energy \eqref{EnergyConsNewton} and
Navier-Stokes equation \eqref{NavierStokesNewton} can be expressed
in terms of this stress-energy tensor as
\begin{equation}
 \partial_\mu  \widehat T^\mu{}_\nu - n \partial_\nu \widetilde\Phi = 0 \ .
\end{equation}

\section{The entropy current}\label{sec:entropy}

The holographic entropy current $J_S^\mu$ is defined by the dual of
the $(d-1)$-dimensional volume form on the time slice on the horizon;
\begin{equation}
 \epsilon_{\mu_1\cdots\mu_d} J_S^{\mu_1} dx^{\mu_2} \wedge \cdots \wedge dx^{\mu_d} .
\end{equation}
The entropy current $J_S^\mu$ can be expressed in terms of the normal vector $n^\mu$ as
\begin{equation}
 J_S^\mu = \frac{\sqrt{h}}{4G} \frac{n^\mu}{n^0} \ , \label{EntropyCurrent}
\end{equation}
where the normal vector at the horizon is given by
\begin{align}
 n_\mu &= \partial_\mu \mathcal S \ , &
 \mathcal S &= r - r_0(x) \ ,
\end{align}
to first order in the derivative expansion.%
\footnote{
The horizon radius has
corrections at higher order in the derivative expansion.
Since these correction terms appear from the second order,
it does not contribute to the entropy current at the first order.
}

Then, the holographic entropy current
for the Lifshitz space-time with $d=4$ and $z=2$ is obtained as
\begin{align}
 J_S^0 &= \frac{1}{4G} r_0^3 \ ,
\\
 J_S^i &= \frac{1}{4G} r_0^3 v^i - \frac{1}{4G}	r_0^2 \partial_i r_0 \ .
\end{align}
Comparing this expression with
\eqref{EnergyCurrent} and \eqref{FluidVariables},
we find that the entropy current satisfies
\begin{align}
 T J_S^\mu = \widehat{\mathcal{E}}^\mu + P \hat v^\mu
 = - \widehat T^\mu{}_\nu \hat v^\nu\ + P \hat v^\mu \ ,
\label{ThermoRel}
\end{align}
where the Hawking temperature $T$ is given by
\begin{equation}
 T = \frac{5}{4\pi} r_0^2  \ ,
\end{equation}
to first order in the derivative expansion.

We can easily check that the entropy current satisfies the second law.
The divergence of the entropy current is calculated as
\begin{align}
 \partial_\mu J_S^\mu
 &= \frac{1}{4G}\partial_t r_0^3 + \frac{1}{4G} v^i \partial_i r_0^3
 + \frac{1}{4G} r_0^3 \partial_i v^i - \frac{1}{12G} \partial^2 r_0^3 \ ,
\end{align}
By using \eqref{Const2S} and \eqref{Const2T}, it becomes
\begin{align}
 \partial_\mu J_S^\mu
 &= \frac{1}{2G} r_0 (\partial_i r_0)^2
 + \frac{1}{8G} r_0 \sigma_{ij}\sigma_{ij}\geq 0 \ ,
\end{align}
That is manifestly non-negative. It therefore satisfies the second law.

The entropy density $s$ is given by
\begin{equation}
 s = J_S^0 = \frac{1}{4G} r_0^3 \ . \label{EntropyDensity}
\end{equation}
The ratio ${\eta/s}$ saturates the KSS bound \cite{Kovtun:2004de},
\begin{equation}
 \frac{\eta}{s} = \frac{1}{4\pi} \ .
\end{equation}

\section{General background gauge field}\label{sec:GeneralA}

So far we have imposed the regularity condition at the horizon $r=r_0$ to the gauge field $A_\mu$ and considered the following ansatz;
\begin{equation}
 A = a(x) \left(r^{5} - r_0^5(x)\right) dt
 - a(x) r^2 dr + \mathcal A_i(x) (dx^i - v^i(x) dt) \ .
\label{RegA}
\end{equation}
We have found in the previous section that this ansatz, in the Newton frame, implies a very special non-zero Newton potential that is velocity dependent
\begin{equation}
 \widetilde\Phi = \mathcal A_t + \frac{1}{2} v^2 = - \frac{1}{2} v^2 \ .
\end{equation}

We will relax here this regularity condition at the horizon because,
as discussed in Appendix~\ref{app:RegA},
even if we do not impose this condition,
the singularity appears only at the past horizon.
Therefore, we can consider a more general background for the boundary theory.

To implement this, we consider a more general ansatz;
\begin{equation}
 A = a(x) \left(r^{5} - r_0^5(x)\right) dt
 - a(x) r^2 dr + \mathcal A_t(x) dt + \mathcal A_i(x) dx^i \ .
\end{equation}
The solution for the correction terms is not modified by this generalization,
\begin{equation}
 A = a(x) \left[\left(r^{5} - r_0^5(x)\right)
 - \frac{1}{3}r^3 \partial_i v^i(x)\right]dt
 - a(x) r^2 dr + \mathcal A_t(x) dt + \mathcal A_i(x) dx^i \ ,
\end{equation}
but the constraint \eqref{ConstS} is modified to
\begin{equation}
 \partial_i r_0^5 = \frac{1}{a} \mathcal F_{i\mu} \hat v^\mu \label{ConstG}
\end{equation}
where
\begin{align}
 \mathcal F &= d\mathcal A \ , &
 \mathcal A &= \mathcal A_t dt + \mathcal A_i dx^i \ .
\end{align}
By identifying as before $\mathcal A$ to the gauge field in the Newton-Cartan in the holographic frame, we find that the gauge field in the Newton frame $ \widetilde{\mathcal A}$ is given by
\begin{equation}
 \mathcal A= \widetilde{\mathcal A} + v^i dx^i - \frac{1}{2} v^2 dt \ ,
 \label{Lif2NC}
\end{equation}
and the constraint \eqref{ConstG} becomes
\begin{equation}
 \partial_i r_0^5 + \frac{1}{a}\partial_t v^i + \frac{1}{a} v^j \partial_j v^i
 = \frac{1}{a}\widetilde{\mathcal F}_{i\mu}\hat v^\mu \ .
\end{equation}
Taking into account the second order terms we obtain
\begin{equation}
 \partial_i P + n \partial_t v^i
 + n v^j \partial_j v^i
 - \partial_j \left(\eta\sigma_{ij} \right) = \widetilde{\mathcal F}_{i\mu} J^\mu \ .
\label{NavierStokesExt}
\end{equation}
where now $ \widetilde{\mathcal F}_{i\mu}$ is a general field strength.
This is in agreement with the non-relativistic Navier-Stokes equation \eqref{NRNavierStokes}
with an external gauge field term,
which contains the Newton potential $\widetilde\Phi$
in $\widetilde F$ as $\widetilde{\mathcal A}_t = \widetilde\Phi$.
The energy conservation \eqref{LifContinuity} does not change but
an appropriate combination with \eqref{NavierStokesExt} gives
the conservation of total energy in the presence of the external field;
\begin{align}
 0 &= \partial_t \left(\mathcal E + \frac{1}{2} n v^2 - n\widetilde\Phi\right)
 + \partial_i \left[\left(\mathcal E + P + \frac{1}{2} n v^2 - n\widetilde\Phi\right)v^i
  - \eta \sigma_{ij} v^j - \kappa \partial_i T \right]
\notag\\&\qquad
+ n \partial_t\widetilde\Phi + n v^j \partial_i \widetilde{\mathcal A}_j\ .
\label{TotalEnergyConsExt}
\end{align}
Now, there are no constraints on the Newton potential
$\widetilde\Phi$ and $\widetilde{\mathcal A}_i$, and hence they are arbitrary.

Next, we return to  the stress-energy tensor which was defined in (\ref{T_hat}).
Including the general gauge field, a readaptation of our previous calculation gives now the following result for the  renormalized stress-energy tensor,
\begin{align}
 \widehat T^0{}_0
 &=
 \frac{1}{8\pi G}  \left(-\frac{3}{2} r_0^5 - \frac{1}{a} v^i \mathcal A_i \right)
  \ ,
\label{T00G}
\\
\widehat T^i{}_0
 &=
 \frac{1}{8\pi G}  \left(- \frac{5}{2} r_0^5 v^i + \frac{1}{2} \partial_i r_0^5 + \frac{1}{a} v^i \mathcal A_t
 + \frac{1}{2} r_0^3 \sigma_{ij} v^j  \right)\ ,
\label{Ti0G}
\\
\widehat T^0{}_i
 &=
 \frac{1}{8\pi G} \frac{1}{a} \mathcal A_i  \ ,
\label{T0iG}
\\
\widehat T^i{}_j
 &=
 \frac{1}{8\pi G} \left( r_0^5 \delta_{ij}
 - \frac{1}{a}(\mathcal A_t + v^k \mathcal A_k) \delta_{ij} - \frac{1}{2} r_0^3 \sigma_{ij}
 + \frac{1}{a} v^i \mathcal A_j \right)
  \ .
\label{TijG}
\end{align}
As already mentioned, this stress-energy tensor includes contributions from the current and gauge field. It is not gauge invariant but is Milne-boost invariant

In this class of theories we can do several redefinitions of the stress-energy tensor preserving its conservation but changing its transformation properties. This will affect the form of the conservation equations but not their physics. We will discuss in the rest of this section two such redefinitions that are interesting.

\subsection{A gauge invariant and Milne-boost invariant stress-energy tensor}

This  is given as
\begin{equation}
 T^\mu{}_\nu =  \widehat T^\mu{}_\nu -J^\mu \mathcal A_\nu + \delta^\mu{}_\nu J^\rho \mathcal A_\rho \ ,
 \label{PhysSE}
\end{equation}
where $J^\mu = n \hat v^\mu$.
The $ T^\mu{}_\nu$ defined above is both gauge invariant and Milne-boost invariant.
In order to see that the above definition gives a Milne-boost invariant,
it is convenient to express $T^\mu{}_\nu$ as follows
\begin{equation}
 T^\mu{}_\nu
 = - \mathcal E \hat v^\mu \tau_\nu
 + P \hat e_a^{\mu} \hat e^a_\nu
 - \eta \sigma_{ab} \hat e_a^{\mu} \hat e^b_\nu
 + \kappa \tau_\nu h^{\mu\rho} \partial_\rho T \ .
\end{equation}
Milne boost invariance is manifest since all terms above are Milne-boost invariant.

We now define again the energy and momentum
\begin{align}
 \mathcal E^\mu
 &= - T^\mu{}_\nu \hat v^\nu \ ,
\label{Edef1}
\\
 \mathcal P_\mu
 &=
  T^\rho{}_\nu \tau_\rho \hat e_a{}^\nu \hat e^a{}_\mu  \ ,
\label{Pdef1}
\\
 \mathcal T^\mu{}_\nu
 &=
 T^\rho{}_\sigma (\hat e^a{}_\rho \, \hat e_a{}^\mu) (\hat e_b{}^\sigma \, \hat e^b{}_\nu) \ .
\label{Tdef1}
\end{align}
in analogy with (\ref{Edef})-(\ref{Tdef}).
With this definition, the energy flow $\mathcal E^\mu$ and stress-energy tensor $\mathcal T^i{}_j$
are the
same to $\widehat{\mathcal E}^\mu$ and $\widehat{\mathcal T}^i{}_j$,
which are defined by $\widehat T^\mu{}_\nu$, \eqref{EnergyCurrent} and \eqref{StressTensor}, respectively.
The momentum density however does not have a contribution from the external gauge fields and vanishes identically
\begin{equation}
 \mathcal P_i = 0  \ .
\end{equation}
This is different from the usual non-relativistic momentum density
but is consistent with the Ward identities of a Milne-invariant theory
\begin{equation}
 \mathcal P_\mu = h_{\mu\nu}J^\nu = 0 \ .
\end{equation}
In fact, the energy flow $\mathcal E^\mu$ is also different from
the standard definition of the energy flow,
but equivalent to the Milne invariant part of the energy flow in \cite{Jensen:2014ama}.

The conservation equations in the Newton-Cartan theory with external gauge field terms
\begin{align}
 \nabla_\mu \mathcal E^\mu &=
 - \frac{1}{2} (\nabla^\mu \hat v^\nu + \nabla^\nu \hat v^\mu) \mathcal T_{\mu\nu}
 + \hat v^\mu \mathcal F_{\mu\nu} J^\nu \ ,
\label{ConsTwF}
\\
 \nabla_\mu \mathcal T^\mu{}_i &= \hat v^\mu \nabla_i \mathcal P_\mu
 - \nabla_\mu (\hat v^\mu \mathcal P_i) + \mathcal F_{i\mu} J^\mu \ ,
\label{ConsSwF}
\end{align}
agree with the constraints from the bulk equations of motion and
become
\begin{align}
 0 &= \partial_t \mathcal E + v^i \partial_i \mathcal E + (\mathcal E + P) \partial_i v^i
 - \frac{1}{2} \eta \sigma_{ij}\sigma_{ij}
 - \partial_i( \kappa \partial_i T)  \ , \label{LifContinuityG}
\\
 0 &= \partial_i P
  - \partial_j \left(\eta\sigma_{ij} \right) - \mathcal F_{i\mu} J^\mu
\ , \label{LifNavierStokesG}
\\
 0 &= \partial_t n + \partial_j (n v^j) \ ,  \label{LifChargeConsG}
\end{align}
where energy density $\mathcal E$, pressure $P$ and particle number density $n$
are defined in \eqref{FluidVariables}.
By using \eqref{Lif2NC}, \eqref{LifNavierStokesG} becomes
\begin{equation}
  \partial_i P + n \partial_t v^i
 + n v^j \partial_j v^i
 - \partial_j \left(\eta\sigma_{ij} \right) = \widetilde{\mathcal F}_{i\mu} J^\mu \ .
\end{equation}
Then, these equations agree with the standard non-relativistic fluid equations
with external sources $\widetilde{\mathcal A}$.
Since the energy flow $\mathcal E^\mu$ and stress tensor $\mathcal T^i{}_j$ are the same
to those defined with $\widehat T^\mu{}_\nu$,
the Lifshitz scaling condition, $z \mathcal E = (d-1) P$ can be expressed in terms of
this stress-energy tensor $T^\mu{}_\nu$ as
\begin{equation}
 z \tau_\mu \hat v^\nu T^\mu{}_\nu + \hat e^a_\mu \hat e_a^\nu T^\mu{}_\nu = 0 \ .
\end{equation}
The thermodynamic relations can be expressed in a similar fashion to \eqref{ThermoRel};
\begin{align}
 T J_S^\mu
 &= \mathcal E^\mu + P \hat v^\mu
\notag\\
 &= - T^\mu{}_\nu \hat v^\nu + P \hat v^\mu
\notag\\
 &= - \widehat T^\mu{}_\nu \hat v^\nu + P\hat v^\mu \ .
\end{align}

\subsection{A gauge invariant stress-energy tensor that is not Milne-boost invariant}

Another possible redefinition of the stress-energy tensor is as follows,
\begin{equation}
 \bar T^\mu{}_\nu =
\widehat T^\mu{}_\nu \hat v^\nu - J^\mu \widetilde A_\nu + \delta^\mu{}_\nu J^\rho \mathcal A_\rho \ .
\label{Tbar}
\end{equation}
Since $\widehat T^\mu{}_\nu$ is Milne-boost invariant and
$\widetilde{\mathcal A}$ is  Milne-boost non-invariant,
this stress-energy tensor is gauge invariant but Milne-boost non-invariant.
By using the identification \eqref{Lif2NC}, this stress-energy tensor is expressed as
\begin{align}
 \bar T^0{}_0
 &=
 - \left(\mathcal E 
 + \frac{1}{2} n v^2 \right) \ ,
\\
 \bar T^i{}_0
 &=
 - \left(\mathcal E + P 
 + \frac{1}{2} n v^2 \right) v^i
 + \eta\sigma_{ij} v^j + \kappa \partial_i T \ ,
\\
 \bar T^0{}_i
 &=
 n v^i 
 \ ,
\\
 \bar T^i{}_j
 &=
 P \delta_{ij} - \eta \sigma_{ij}
 + n v^i v^j 
 \ .
\end{align}
This stress-energy tensor takes the same form to
that for the standard non-relativistic fluids.%
\footnote{Here, the stress-energy tensor consists of the energy flow, momentum density and stress tensor.
Sometimes, the stress-energy tensor is constructed by using the mass flow instead of the energy flow
(see, for example \cite{Jensen:2014ama}).
Our stress-energy tensor is different from the stress-energy tensor which contains mass flow.
}
In order to see that $\bar T^\mu{}_\nu$ is not invariant under a Milne boost, explicitly,
we rewrite it as
\begin{equation}
 T^\mu{}_\nu
 = - \left(\mathcal E + \frac{1}{2}n \bar h_{\rho\sigma} \hat v^\rho \hat v^\sigma\right)
 \hat v^\mu \tau_\nu
 + P \hat e_a^{\mu} \hat e^a_\nu
 + n \hat v^\mu \bar h_{\nu\rho} \hat v^\rho
 - \eta \sigma_{ab} \hat e_a^{\mu} \hat e^b_\nu
 + \kappa \tau_\nu h^{\mu\rho} \partial_\rho T \ .
\end{equation}
This expression contains the Milne-boost non-invariant, $\bar h_{\mu\nu}$.
Because of these terms, $\bar T^\mu{}_\nu$ is not invariant under the Milne boost.
The energy current $\bar{\mathcal E}^\mu$, momentum density $\bar{\mathcal P}_\mu$,
and stress tensor $\bar{\mathcal T}^i{}_j$ are defined by
\begin{align}
 \bar{\mathcal E}^\mu
 &= - \bar T^\mu{}_\nu \bar v^\nu \ ,
\label{PhysE}
\\
 \bar{\mathcal P}_\mu
 &=
 \bar T^\rho{}_\nu \tau_\rho \bar e_a^\nu \bar e^a_\mu \ ,
\label{PhysP}
\\
 \bar{\mathcal T}^\mu{}_\nu
 &=
 \bar T^\rho{}_\sigma (\bar e^a_\rho \, \bar e_a^\mu) (\bar e_b^\sigma \, \bar e^b_\nu) \ .
\label{PhysT}
\end{align}
where $\bar e^a_\mu = \bar e_a^\mu = \mathrm{diag}(0,1,1,1)$, or equivalently,
\begin{align}
 \bar{\mathcal E}^\mu
 &= - T^\mu{}_0 \ ,
&
 \bar{\mathcal P}_i
 &=
 \bar T^0{}_i  \ ,
&
 \bar{\mathcal T}^i{}_j
 &=
 \bar T^i{}_j \ .
\end{align}
Now the energy flow, momentum density and stress tensor are similar to
those for a  standard non-relativistic fluid.
For example, the momentum density is given by the standard form, $\bar{\mathcal P}_i = nv^i$,
which is related to the velocity field $v^i$.

The conservation law becomes
\begin{align}
 \nabla_\mu \bar{\mathcal E}^\mu &=
 - \frac{1}{2} (\nabla^\mu \bar v^\nu + \nabla^\nu \bar v^\mu) \bar{\mathcal T}_{\mu\nu}
 + \bar v^\mu \widetilde{\mathcal F}_{\mu\nu} J^\nu \ ,
\label{ConsTwFwbar}
\\
 \nabla_\mu \bar{\mathcal T}^\mu{}_i &= \bar v^\mu \nabla_i \bar{\mathcal P}_\mu
 - \nabla_\mu (\bar v^\mu \bar{\mathcal P}_i) + \widetilde{\mathcal F}_{i\mu} J^\mu \ .
\label{ConsSwFwbar}
\end{align}
In this case, the conservation law can be expressed in terms of the stress-energy tensor
$\bar T^\mu{}_\nu$ as
\begin{align}
 \partial_\mu \bar T^\mu{}_\nu - J^\mu \widetilde{\mathcal F}_{\mu\nu} &= 0 \ , \\
 \partial_\mu J^\mu &= 0 \ .
\end{align}
The conservation of total energy and the Navier-Stokes equation are expressed as
\begin{align}
 0 &= \partial_t \left(\mathcal E + \frac{1}{2} n v^2 - n\widetilde\Phi\right)
 + \partial_i \left[\left(\mathcal E + P + \frac{1}{2} n v^2 - n\widetilde\Phi\right)v^i
  - \eta \sigma_{ij} v^j - \kappa \partial_i T \right]
\notag\\&\qquad
+ n \partial_t\widetilde\Phi + n v^i \partial_t \widetilde{\mathcal A}_i\ , \\
 0 &= \partial_i P + n \partial_t v^i
 + n v^j \partial_j v^i
 - \partial_j \left(\eta\sigma_{ij} \right) - \widetilde{\mathcal F}_{i\mu} J^\mu \ ,
\end{align}
where $\widetilde{\mathcal F}$ contains the Newton potential $\widetilde\Phi=\widetilde{\mathcal A}_t$.

For $z=2$, the Lifshitz scaling invariant condition is expressed in terms of $\bar T^\mu{}_\nu$ as
\begin{equation}
 z \bar T^0{}_0 + \bar T^i{}_i = 0 \ .
\end{equation}
however, this expression is valid only for $z=2$.
For general $z$, which we will discuss in Section~\ref{sec:general},
the Lifshitz scaling invariant condition is given by
\begin{equation}
 z(\bar T^0{}_0 + J^0 \widetilde{\mathcal A}_0 - J^0 \mathcal A_0)
 + \bar T^i{}_i + J^i \widetilde{\mathcal A}_i - J^i \mathcal A_i = 0 \ .
\label{LifWardIdNewton}
\end{equation}
The thermodynamic relation is expressed as
\begin{align}
 T J_S^\mu
 &= - (\bar T^\mu{}_\nu + J^\mu \widetilde A_\nu) \hat v^\nu + \hat\mu J^\mu + P \hat v^\mu
\notag\\
 &= - \widehat T^\mu{}_\nu \hat v^\nu + P \hat v^\mu \ ,
\end{align}
where the chemical potential $\hat\mu$ is given by
\begin{equation}
 \hat\mu = \hat v^\mu \mathcal A_\mu 
\ .
\end{equation}

Fluids in the Newton-Cartan theory were studied in \cite{Jensen:2014ama}.
The relation between the variables in this paper and those in \cite{Jensen:2014ama}
are presented in Appendix~\ref{app:notation}.
The fluid obtained in this paper is the same to that in \cite{Jensen:2014ama},
but some transport coefficients are absent here.
There are 6 transport coefficients in \cite{Jensen:2014ama} and
3 of them are related to parity-odd terms. Such terms are not present here as we ignored all potential CP-odd terms in the bulk effective action. Adding the gravitational, gauge and mixed Chern-Simons terms we expect that such CP-odd terms will be generated but we leave this for future work.

The bulk viscosity vanishes here because of the Lifshitz scaling symmetry.
Finally the remaining two  transport coefficients,
the shear viscosity and heat conductivity appear in our model and take concrete values associated to the bulk theory in question.
Furthermore, we study only the flat background for
the geometry on the boundary, $\tau_\mu$ and $\hat e^\mu_a$, and hence,
related quantities such as torsion, do not appear in this paper.

\section{The case of general $z$}\label{sec:general}

So far, we have focused on the case of $z=2$.
In this section, we consider the hydrodynamic ansatz for $z>1$.

As in the case of $z=2$, we first replace the parameters by the slowly varying functions.
For general $z$, we start from the following ansatz;
\begin{align}
 ds^2 &= - (r^{2z} f - v^2(x) r^2) dt^2 + 2 r^{z-1} dt dr - 2 r^2 v^i(x) dt\, dx^i + r^2 (dx^i)^2
\label{BGmetricZ}
\\
 f &= 1 - \frac{r_0^{z+3}(x)}{r^{z+3}}
\\
 A &= \left[a(x) \left(r^{z+3} - r_0^{z+3}(x)\right) - \mathcal A_i(x) v^i(x)\right]dt
 - a(x) r^2 dr + \mathcal A_i(x) dx^i,
\\
 e^{\lambda\phi} &= \mu (x) r^{-6} ,
\end{align}
and calculate the correction terms by using the derivative expansion.

As for $z=2$, we introduce the correction terms and
solve the linear differential equations.
The integration constant can be fixed by the requirement for the asymptotic behavior and
the regularity at the horizon.
The first order solution becomes
\begin{align}
 ds^2 &= - r^{2z} f dt^2 + 2 r^{z-1} dt dr
 + r^2 (dx^i - v^i dt)^2
\notag\\&\quad
 + \frac{2}{3}r^z \partial_i v^i dt^2
 + 2 F_3(r) \partial_i r_0 dt (dx^i - v^i dt)
 - r^2 F_1(r) \sigma_{ij} (dx^i - v^i dt) (dx^j - v^j dt)
\label{SolZ}
\end{align}
and
\begin{equation}
 A = a \left(r^{z+3} - r_0^{z+3} - \frac{1}{3} \partial_i v^i\right) dt
 - a r^2 dr + F_2(r) \partial_i r_0 (dx^i - v^i dt) \ ,
\end{equation}
where
\begin{align}
 F_1(r)
 &=
 \int dr \frac{r^3-r_0^3}{r(r^{z+3}-r_0^{z+3})}  \ ,
\\
 F_2(r)
 &=
 \left( 2(z-1) r^{z+3} - (z-5) r_0^{z+3}\right)
\notag\\&\quad\times
 \int dr \frac{(z+3) a r^2  r_0^{z+2}
 \left[10 (z-1) r^{z+3} r_0^2 + z(z+3) r^5 r_0^z - (z-5)(z-2) r^{z+3} \right]}
 {2(z-1) (r^{z+3}-r_0^{z+3})[2(z-1) r^{z+3} - (z-5) r_0^{z+3}]^2} \ ,
\\
 F_3(r)
 &=
 - \int \frac{dr}{r^{6-z}}\frac{2(z-1)}{a} F_2(r) \ ,
\end{align}
where the integration constants are determined by
the asymptotic behavior at the boundary, $r\to\infty$,
\begin{align}
 F_1(r) &= \mathcal O(r^{-z}) \ , &
 F_2(r) &= \mathcal O(r^{2-2z}) \ , &
 F_3(r) &= \mathcal O(r^{-z-1}) \ , &
\end{align}
for $1<z \leq 2$. For $z>2$, they have different asymptotic behavior but
the integration constants can be taken to be the analytic continuation of those for $1<z \leq 2$.

The functions $r_0$, $v^i$, $a$ and $\mathcal A_i$
in the  solution must satisfy the following constraints;
\begin{align}
 0 &= \partial_t a + v^i \partial_i a - a \partial_i v^i  ,
\\
 0 &= \partial_t r_0 + v^i \partial_i r_0 + \frac13 r_0 \partial_i v^i  ,
\\
 0 &= \partial_t \mathcal A_i + v^j \partial_j \mathcal A_i + \mathcal A_j \partial_i v^j
 + \frac{z (z+3)}{2 (z-1)} r_0^{z+2} a \partial_i r_0 \ .
\end{align}
These constraints agree with the leading order fluid equations.

In order to obtain regular results,
we introduce the following counter terms;
\begin{equation}
 S_\text{ct} =
 \frac{1}{16\pi G} \int d^4 x \sqrt{-\gamma}
 \left[
 - (5+z) + \frac{z+d-1}{2} \, e^{\lambda\phi} \gamma^{\mu\nu} A_\mu A_\nu
 \right] \ .
\end{equation}
Then, the stress-energy tensor is calculated from $\mathcal O(r^{z+3})$ terms as
\begin{align}
 \widehat T^0{}_0
 &=
 \frac{1}{8\pi G}\left(- \frac{3}{2} r_0^{z+3} - \frac{z-1}{a} v^i \mathcal A_i \right)
 \ ,
\\
 \widehat T^i{}_0
 &=
 \frac{1}{8\pi G} \left(-\frac{z+3}{2} r_0^{z+3} v^i + \frac{z(z+3)}{4(z-1)} r_0^{2z} \partial_i r_0
 - \frac{z-1}{a} v^i v^j \mathcal A_j
 + \frac{1}{2} r_0^3 \sigma_{ij} v^j  \right) \ ,
\\
 \widehat T^0{}_i
 &=
 \frac{1}{8\pi G}\frac{z-1}{a} \mathcal A_i \ ,
\\
 \widehat T^i{}_j
 &=
 \frac{1}{8\pi G} \left(\frac{z}{2} r_0^{z+3} \delta_{ij} - \frac{1}{2} r_0^3 \sigma_{ij}
 + \frac{z-1}{a} v^i \mathcal A_j \right)
 \ .
\end{align}

{}From the above expression of the stress-energy tensor,
we identify the fluid variables as
\begin{align}
 \mathcal E &= \frac{3}{16\pi G} r_0^{z+3} \ , &
 n &= \frac{z-1}{16\pi G a} \ , &
 P &= \frac{z}{16\pi G} r_0^{z+3} \ .
\label{FluidVariablesZ}
\end{align}
Here, the energy density and pressure satisfies the scaling invariant condition
\begin{equation}
 z\mathcal E = (d-1) P .
\end{equation}
The energy flow, momentum density and stress tensor are expressed as
\begin{align}
 \mathcal E^0 &= \mathcal E \ ,
\\
 \mathcal E^i &= \mathcal E v^i - \kappa \partial_i T \ ,
\\
 \mathcal P_i &= n \mathcal A_i \ ,
\\
 \mathcal T^i{}_j &= P \delta_{ij} - \eta \sigma_{ij} \ ,
\end{align}
where
\begin{align}
 \kappa &= \frac{1}{8(z-1)G} r_0^{z+1} \ ,
&
 \eta &= \frac{1}{16\pi G} r_0^3 \ .
\label{TransZ}
\end{align}
The fluid variables and transport coefficients can be expressed in terms of the temperature
\begin{equation}
 T = \frac{z+3}{4\pi} r_0^z \ ,
\end{equation}
as
\begin{align}
 \mathcal E &= \frac{3}{16\pi G} \left(\frac{4\pi}{z+3}T\right)^{\frac{z+3}{z}} \ , &
 P &= \frac{z}{16\pi G} \left(\frac{4\pi}{z+3}T\right)^{\frac{z+3}{z}} \ ,
\end{align}
and
\begin{align}
 \kappa &= \frac{1}{8(z-1)G} \left(\frac{4\pi}{z+3}T\right)^{\frac{z+1}{z}} \ ,
&
 \eta &= \frac{1}{16\pi G} \left(\frac{4\pi}{z+3}T\right)^{\frac{3}{z}} \ .
\label{kal}\end{align}
The scaling dimension under the Lifshitz scaling would be
\begin{align}
 [\mathcal E] &= z+3 \ , &
 [P] &= z+3 \ , \\
 [n] &= 3 \ , &
 [T] &= z \ , \\
 [v^i] &= z-1 \ , &
 [\mathcal A_i] &= 1 \ , \\
 [\kappa] &= z+1 \ , &
 [\eta] &= 3 \ .
\end{align}
Note that all dimensions above are the canonical dimensions we expect in a Lifshitz-invariant theory.
Moreover, as shown in \cite{Jensen:2014ama}, the heat conductivity $\kappa$ is related to the standard DC conductivity $\sigma$ by
\be
 \kappa = {(\mathcal E+P)^2\over n^2~ T}\sigma
 \ee
This relation is compatible with $[\s]=d-2$.  Substituting $\sigma\sim T^{-{5\over z}}$ from (\ref{18}) and $\mathcal E\sim P\sim T^{z+3\over z}$, $n\sim T^0$ found above we find $\kappa\sim T^{{z+1\over z}}$ compatible with (\ref{kal}) above.

The conservation law can be understood as the following fluid equations,
\begin{align}
 0 &= \partial_t \mathcal E + v^i \partial_i \mathcal E + (\mathcal E + P) \partial_i v^i
 - \frac{1}{2} \eta \sigma_{ij}\sigma_{ij}
 - \partial_i \left( \kappa \partial_i T \right)
\ , \label{LifContinuityZ}
\\
 0 &= \partial_i P + n \partial_t \mathcal A_i
 + n v^j \partial_j \mathcal A_i + n \mathcal A_j \partial_i v^j
  - \partial_j \left(\eta\sigma_{ij}
 \right)
\ , \label{LifNavierStokesZ}
\\
 0 &= \partial_t n + \partial_j (n v^j) \ .
\end{align}
These equations have the same form to \eqref{LifContinuity}-\eqref{LifChargeCons} and
the differences appear only in
the fluid variables \eqref{FluidVariablesZ} and transport coefficients \eqref{TransZ}.
As we have discussed in Section~\ref{NC},
these equations can be identified to the standard fluid equations
by identifying $\mathcal A$ as the gauge field in the Newton-Cartan theory
in the holographic frame $\bar v^\mu = \hat v^\mu$.

It should be noted that a constant with Lifshitz scaling dimension $2-z$
should be introduced to the Milne boost, since
the scaling dimensions of $\mathcal A_i$ and $v^i$ are different.
Here, we refer to it as $m$ and then \eqref{gaugeNC} is modified as
\begin{equation}
 \widetilde{\mathcal A} = \mathcal A - m \left(v^i dx^i - \frac{1}{2} v^2 dt \right) \ .
\end{equation}
This implies that $\frac{1}{m}\mathcal A$ corresponds
to the gauge field in Newton-Cartan theory, $B$, in the holographic frame.
The fluid equations in the Newton frame are expressed as
\begin{align}
 0 &= \partial_t \mathcal E + v^i \partial_i \mathcal E + (\mathcal E + P) \partial_i v^i
 - \frac{1}{2} \eta \sigma_{ij}\sigma_{ij}
 - \partial_i( \kappa \partial_i T)  \ ,
\\
 0 &=   \partial_i P + \rho \partial_t v^i
 + \rho v^j \partial_j v^i
 - \partial_j \left(\eta\sigma_{ij} \right) - m {\mathcal B}_{i\mu} J^\mu \ ,
\\
 0 &= \partial_t n + \partial_j (n v^j) \ .
\end{align}
Here, $\mathcal B_{\mu\nu} = \frac{1}{m}\widetilde{\mathcal F}_{\mu\nu}$ is the two-form in the Newton-Cartan theory in the Newton frame, and $\rho = mn$.
Since $m$ is the coupling constant for gravitational potential,
it can be interpreted as the mass per particle,
and $\rho = mn$ is the mass density,
which has a different scaling dimension to particle number density for general $z$.
In the Newton frame, the constant $m$ should scale
with the scaling dimension $2-z$ under the Lifshitz scaling transformation,
in addition to the ordinary Lifshitz scaling transformation.
This is an analog of the generalized conformal symmetry in which
the coupling constant also scales under the scaling symmetry.
In fact, the Lifshitz scaling condition \eqref{LifWardIdNewton} is expressed as
\begin{equation}
 z\bar T^0{}_0 + \bar T^i{}_i + \frac{z-2}{2} m n v^2 = 0 \ .
\end{equation}
The last term implies the additional transformation of $m$.

For general $z$, the entropy current is calculated as
\begin{align}
 J_S^0 &= \frac{1}{4G} r_0^3 \ ,
\\
 J_S^i &= \frac{1}{4G} r_0^3 v^i - \frac{z}{8(z-1)G} r_0^z \partial_i r_0 \ .
\end{align}
They satisfy the the thermodynamic relation;
\begin{align}
 T J_S^\mu = {\mathcal{E}}^\mu + P \hat v^\mu
 = - \widehat T^\mu{}_\nu \hat v^\nu + P \hat v^\mu \ .
\end{align}
In terms of the temperature, entropy density is expressed as
\begin{equation}
 s = J^0_S = \frac{1}{4G} \left(\frac{4\pi}{z+3}T\right)^{\frac{3}{z}} \ .
\end{equation}

\section{Results, interpretation and outlook}\label{sec:conclusion}

In this paper, we have investigated the fluid/gravity correspondence for Lifshitz invariant theories.
We have considered the Einstein-Maxwell-Dilaton theory
which has the Lifshitz geometry as a solution. The gauge field and dilaton break however this scaling symmetry mildly and this breaking is characterized by a non-trivial conduction exponent $\psi$.

The geometry which describes hydrodynamics is constructed
from the black hole geometry.
We have used the Eddington-Finkelstein coordinates to
impose the regularity condition at the future horizon,
and we have then introduced a Galilean boost.
We have replaced the boost parameter and horizon radius by
slowly varying functions of the space and time coordinates. We have done the same for the parameters appearing in the gauge field and the dilaton.
The solution is derived to first order but
the constraints are calculated to the second order.

The relation between the bulk constraint equations and
the conservation laws on the boundary implies that
the boundary theory satisfies the conservation laws of a Newton-Cartan theory.
The conservation laws have the form of  fluid equations but
in a non-trivial Newtonian potential,
if we impose the ordinary regularity condition of the gauge field, $A_t=0$ at the horizon.
However, the geometry which describes the fluid  is generically not regular at the past horizon,
even if the regularity condition at the past horizon is imposed on the gauge field.
Imposing the regularity condition on the gauge field only at the future horizon,
we obtain the fluid equations with arbitrary Newton-Cartan gauge field,
which contains an arbitrary Newton potential.

The fluid has the following properties;

\begin{itemize}
\item
The stress-energy tensor has a form similar to that of non-relativistic fluids, and is expressed in terms of the fluid variables:
the velocity field $v^i$, the energy density $\mathcal E$, the pressure $P$ and the charge density $n$. It also contains the external gauge field $\mathcal A$.
At first order, it has as transport coefficients the
thermal conductivity $\kappa$ and shear viscosity $\eta$ while the bulk viscosity is zero due to the Lifshitz scaling symmetry.
All the above variables and transport coefficients are functions  of
temperature $T$.

\item
The (particle number) density appears associated to the external gauge field.
The stress-energy tensor calculated directly from the bulk solution agrees with that of non-relativistic fluids
except for the terms where the  density appears.
The  terms which contain the density are different from those of a  non-relativistic fluid.
In particular, the momentum density $T^0{}_i$ is proportional to $\mathcal A_i$
and vanishes for $\mathcal A_i=0$.

\item
The conservation law of the stress-energy tensor is not given in the form of its covariant derivative,
but is that appropriate to a boundary Newton-Cartan theory.
It is expressed in terms of the energy flow $\mathcal E^\mu$,
momentum density $\mathcal P_i$ and stress tensor $\mathcal T^i{}_j$.
The stress-energy tensor satisfies the Ward identity of the scale invariance
$z\mathcal E = (d-1) P$ without any modification.
In terms of the fluid variables, the conservation law takes
the form of the fluid equations: the energy conservation, continuity equation and
the Navier-Stokes equation.

\item
The fluid equations are similar to the non-relativistic fluid equations.
The continuity equation and energy conservation equation agree with
that for standard non-relativistic fluids.
However, the Navier-Stokes equation is different from that for ordinary non-relativistic fluids.
It does not contain the velocity field except the terms appearing  in the shear tensor.
It contains the coupling to the external gauge field instead. A related property of the associated stress tensor is that it is gauge non-invariant but Milne-boost invariant.

\item

The absence of the velocity fields in our Navier-Stokes equation can be explained as follows.
For an ordinary fluid, the pressure is comparable to the non-relativistic energy which does not include the mass energy, and hence it is much smaller than the relativistic energy density.
In our case, the pressure is of the same order as the energy density
because of the Lifshitz scaling symmetry.
For this to happen the fluid equations must be different and this is what we find, namely, the contribution from pressure is much larger than that in ordinary fluids, and then,
the terms with velocity fields becomes negligible compared to the pressure.

This is the reason that some terms with velocity fields in the Navier-Stokes equation are absent.
These terms are replaced by the external source
which is identified with the gauge field in the Newton-Cartan theory.

\item We may do some redefinitions of the stress-energy tensor and
by identifying the gauge field $\mathcal A$ to the gauge field in Newton-Cartan theory,
which takes the form of
${\mathcal A} = \widetilde{\mathcal A} + v^i dx^i - \frac{1}{2}v^2 dt$
our Navier-Stokes equation agrees with the standard  non-relativistic Navier-Stokes equation.
In such a case the stress-energy tensor is gauge invariant but Milne-boost non-invariant.
Finally there is a definition of the stress-energy tensor that is both Milne-boost and gauge invariant.

\item
Since the gauge field in the Newton-Cartan theory is a generalization of
the Newtonian gravity theory, a general external gauge field gives
a fluid in a non-trivial gravitational potential.
The Newton potential appears in the gauge field as $\widetilde\Phi = \widetilde{\mathcal A}_t$.
The Navier-Stokes equation we obtain agrees  with the ordinary Navier-Stokes equation in the presence of an external (gravitational) force.
The gauge field does not contribute to
the conservation of (internal) energy density $\mathcal E$.
The conservation of total energy can be obtained from the conservation of $\mathcal E$ and
the Navier-Stokes equation and agrees with that for ordinary non-relativistic fluid in
a non-trivial Newton potential.

\item
The entropy density is defined in terms of the horizon area and
satisfies the local thermodynamic relation with
energy density and pressure.
The divergence of the entropy current is non-negative,
which is consistent with the second law.

\item
The form of the fluid equations is independent of the Lifshitz exponent $z$
as well as of the conduction exponent $\psi$. This dependence appears at first order, inside the various state functions and therefore only in the constitutive relations.

\end{itemize}

There are several obvious interesting questions that remain unanswered by our work. The first is the extension of our results to hydrodynamics in the presence of  hyperscaling violation in the metric ($\theta\not=0$). This is under way. A naive guess would be that the hydrodynamics would be a dimensional reduction of the one found here, along the lines described in \cite{gk1,gs}.
In particular in \cite{gk1} it was shown that Lifshitz solutions with hyperscaling violation can be obtained as suitable dimensional reductions of higher-dimensional Lifshitz invariant theories without hyperscaling violation.
The associated reduction of the hydrodynamics will provide equations similar to the ones here but with a non-zero bulk viscosity. This needs to be verified.

A further extension involves Lifshitz geometries with broken U(1) symmetry.
This is actively pursued in \cite{ho}.

An interesting question in relation to the above is: what is the appropriate hydrodynamics for QFTs that are RG Flows that interpolate between relativistic and non-relativistic theories.
To motivate the answer to this question, we consider first non-Lorentz invariant (but rotationally invariant) flows between Lorentz invariant fixed points\footnote{The fact that the speed off light can vary on branes was pointed out first in \cite{kk}.}, \cite{tw}, but where the velocity of light in the IR is different for that in the UV. In such a case, the hydrodynamics of this theory, is relativistic, but with a speed of light that is {\em temperature dependent}.

This example suggests that in an (Lorentz-violating) RG flow from a CFT (with an unbroken U(1) symmetry that is used to drive the breaking of Lorentz invariance) to an IR non-relativistic scaling (rotational invariant) geometry at an arbitrary temperature, the hydrodynamics will be again of the relativistic form (but with a general equation of state) and with a speed of light $c(T)$ that is again temperature dependent. In the IR, $c(T\to 0)=\infty$ and the hydrodynamics reduces to the one found here with the U(1) symmetry becoming the mass-related symmetry. This is nothing else than the standard non-relativistic limit\footnote{This is expected to happen along the lines presented in \cite{kj} although this needs to be verified.} of the relativistic hydrodynamics while all thermodynamic functions and transport coefficients are smooth functions of $T$ (if no phase transition exists at finite $T$). Otherwise they follow the standard behavior at phase transitions.

A more general breaking of Lorentz invariance during a RG flow must involve higher form fields of tensors in the bulk, or a multitude of vector fields and the details of the RG flow become complicated. It is important that such flows are analyzed as they hold the key to understanding general non-relativistic flows as well as generalized hydrodynamics of the associated theories.

Finally a more detailed study of the above issues in the absence of U(1) symmetry is necessary.

\section*{Acknowledgements}
\addcontentsline{toc}{section}{Acknowledgements}

We would like to thank A.~Mukhopadhyay for discussions. We especially thank J.~Hartong and N.~Obers for participating in early stages of this work and for extensive and illuminating discussions on the topics presented here.

This work was supported in part by European Union's Seventh Framework Programme
under grant agreements (FP7-REGPOT-2012-2013-1) no 316165, the EU program ``Thales'' MIS 375734
and was also cofinanced by the European Union (European Social Fund, ESF) and Greek national funds through
the Operational Program ``Education and Lifelong Learning'' of the National Strategic
Reference Framework (NSRF) under ``Funding of proposals that have received
a positive evaluation in the 3rd and 4th Call of ERC Grant Schemes.''

\newpage
\appendix

\renewcommand{\theequation}{\thesection.\arabic{equation}}
\addcontentsline{toc}{section}{Appendices}
\section*{APPENDIX}

\section{Notations}\label{app:notation}

Since in the topic treated in this paper there are a lots of variables involved and many  redefinitions,  we present here a list of the variables and their definitions with comments when necessary. We also  present a translation dictionary to  the variables used by Jensen, \cite{Jensen:2014ama} and Hartong et al., \cite{{Hartong:2015wxa}}.
In Table~\ref{table1}, we present the relation between the variables
in this paper and those in \cite{Jensen:2014ama} and \cite{Hartong:2015wxa}.
In Table~\ref{table2}, we present
the correspondence of the fluid variables in this paper and those in \cite{Jensen:2014ama}.

\bigskip
\centerline{\bf Variables defined on the  gravity side}
\bigskip

\begin{itemize}

\item $d$ is the dimension of the space-time boundary, and the dimension of the boundary QFT.

\item
$v^i$: Boost parameter introduced into the (static) black hole geometry in \eqref{BoostedBH}.

\item
$r_0$: the horizon radius which is defined in \eqref{WarpF}.

\item
$a$: the coefficient of the $r^{z+d-1}$ term of the gauge field, introduced in \eqref{PureA}.
Note that $a \neq a_\mu dx^\mu$.

\item
$\mathcal A_\mu$: the constant part of the gauge field, which is defined in \eqref{BGgauge}.
This corresponds to the Milne boost-invariant gauge field
in the Newton-Cartan theory $B_{inv}$, or equivalently,
$B$ in the holographic frame $\bar v^\mu = \hat v^\mu$.
In this paper, it is sometimes expressed as the 1-form $\mathcal A = \mathcal A_\mu dx^\mu$. $\mathcal A_\mu$ eventually becomes $x^{\mu}$ dependent in the hydrodynamic ansatz.

\item
$\hat v^\mu$: the timelike inverse vielbein on the boundary which is defined up to the factor
$r^{-2z}f^{-1}$, or equivalently, defined in \eqref{InvIndm} and \eqref{vh}
and given by $\hat v^\mu = (1,v^i)$ in this paper.
This corresponds to the velocity vector field of the fluid.
The holographic frame of the boundary Newton-Cartan geometry is defined by $\bar v^\mu = \hat v^\mu$.

\item
$\tau_\mu$: the timelike vielbein on the boundary which is defined up to the factor of
$r^{2z}f$, or equivalently, defined in \eqref{Indm} and \eqref{vl}
and given by $\tau_\mu = (1,0)$ in our solution.
This corresponds to the timelike unit normal which defines the time direction in the Newton-Cartan theory.
It is automatically invariant under the Milne boost.

\item
$\hat e^a_\mu$: the spacelike vielbein on the boundary which is defined up to the factor of $r^{2}$, or equivalently, defined in \eqref{Indm} and \eqref{vl},
and given by $e^a_\mu dx^\mu = dx^a - v^a dt$ in this paper.
This is invariant under the Milne boost and
equals to the spacelike vielbein in the Newton-Cartan theory
if we take the holographic frame $\bar v^\mu = \hat v^\mu$.

\item
$\hat e_a^\mu$: the spacelike inverse vielbein on the boundary which is defined up to the factor of
$r^{-2}$, or equivalently, defined in \eqref{InvIndm} and \eqref{vh}, and
given by $e_a^\mu \partial_\mu = \partial_a$ in our model.
This corresponds to the spacelike inverse vielbein.
It is automatically invariant under the Milne boost.

\end{itemize}

\bigskip

\centerline{\bf Variables in the (boundary) Newton-Cartan theory}

\bigskip

\begin{itemize}
\item
$\bar v^\mu$: the timelike inverse vielbein in Newton-Cartan theory.
This notation is introduced above \eqref{tu} to define the Galilei connection.
The timelike inverse vielbein is not invariant under the Milne boost but it is covariant.
In the literature it is sometimes called  ``velocity'' but must be distinguished from the velocity of the fluid.

\item
$\bar h_{\mu\nu}$: the induced covariant metric on the time-slice. It is defined by \eqref{defhbar}
and given by $\bar h_{\mu\nu} = \mathrm{diag}(0,1,1,1)$ for a 4-dim space-time in the Newton frame.
It is not invariant under the Milne boost but it is covariant.

\item
$\bar e^a_\mu$: the spacelike vielbein, which is introduced in \eqref{PhysP}-\eqref{PhysT}.
It is given by $\bar e^a_\mu = \mathrm{diag}(0,1,1,1)$
for a 4-dim space-time. It satisfies $\bar h_{\mu\nu} = \bar e^a_\mu \bar e^a_\nu$.

\item
$\bar e_a^\mu$: the spacelike vielbein, which is introduced in \eqref{PhysP}-\eqref{PhysT}.
It is given by $\bar e_a^\mu = \mathrm{diag}(0,1,1,1)$ for a 4-dim space-time in the Newton frame.
Since it is invariant under a Milne boost, it is equal to $\hat e_a^\mu$.

\item
$B_\mu$: the gauge field
in the Newton-Cartan theory.
This notation is introduced below \eqref{NewtonCond}.
It is not invariant under the Milne boost but it is covariant.

\item
$\widetilde{\mathcal A}_\mu$: the gauge field
in the Newton-Cartan theory in Newton frame $\bar v^\mu = (1,\vec 0)$.
This notation is introduced in \eqref{gaugeNC}.
It is not invariant under the Milne boost but it is covariant.

\item
$\widehat B$: a Milne boost invariant combination for the gauge field.
It is defined in \eqref{new1}, with arbitrary but appropriately normalized
Milne-invariant vector $X^\mu$.

\item
$B_{inv}$: the Milne boost invariant combination $\widehat B$ for $X^\mu = \hat v^\mu$.
It is defined in \eqref{new12}. In terms of the gauge field
in the Newton frame $\widetilde{\mathcal A}$, it is expressed as
$B_{inv} = \widetilde{\mathcal A} + v^i dx^i - \frac{1}{2} v^2 dt$.
It also equals to the gauge field in the holographic frame $\mathcal A$.

\end{itemize}

\begin{table}
\renewcommand{\arraystretch}{1.2}
\begin{center}
\begin{tabular}{|l|c|c|c|c|c|c|c|c|c|c|c|c|c|c|}
\hline
This paper&
$\tau_\mu $ &
$\bar v^\mu  $ &
$\bar h_{\mu\nu} $ &
$\widetilde{\mathcal A}_\mu $ &
$\hat v^\mu $ &
$\hat e^a_\mu$ &
${\mathcal A}_\mu $ &
$n $ &
$\widetilde\Phi$ &
$v^i$
\\\hline
Jensen \cite{Jensen:2014ama} &
$ n_\mu$ &
$ v^\mu$ &
$ h_{\mu\nu}$ &
$ A_\mu$ &
$ u^\mu$ &
-- &
$ \tilde A_\mu$ &
$ \rho$ &
$A_t$ &
$u^i$
\\\hline
HKO \cite{Hartong:2015wxa} &
$\tau_\mu $ &
$- v^\mu  $ &
$h_{\mu\nu} $ &
$- M_\mu$ &
$- \hat v^\mu $ &
$\hat e^a_\mu$ &
$(-\widetilde\Phi,0)$ &
$T^0$ &
$-M_t$ &
$M_i$
\\\hline
\end{tabular}
\caption{
Relation between the variables in this paper and those in \cite{Jensen:2014ama} and \cite{Hartong:2015wxa}.
Correspondence to \cite{Hartong:2015wxa} is read off from the relation of
the metric and gauge field in gravity side.
In \cite{Hartong:2015wxa}, $\hat v^\mu$ is defined such that
spacial components of the associated Milne invariant gauge field vanish.
We can also define such velocity field in our notation,
$\hat v^\mu - h^{\mu\nu} \widehat{\mathcal A}_\nu$.
If we identify this combination to $-\hat v^\mu$ in \cite{Hartong:2015wxa},
we obtain another correspondence.
In this case, no variables in \cite{Hartong:2015wxa} correspond to
the Milne boost invariants in this paper.
}\label{table1}
\end{center}
\renewcommand{\arraystretch}{1}
\end{table}

\bigskip
\centerline{\bf VEVs and fluid variables}
\bigskip

\begin{itemize}
\item
$T_\text{(nr)}{}^\mu{}_\nu$: the stress-energy tensor without counter terms on $dr=0$ (near-boundary)surface.
To be precise, the stress-energy tensor is the coefficient of $\mathcal O(r^{-5})$ term of this tensor
($\mathcal O(r^{-z-3})$ for general $z$).
It appears in \eqref{nrT00}-\eqref{nrTij}

\item
$T_r^\mu{}_\nu$: the renormalized stress-energy tensor on $dr=0$ (near-boundary) surface
or its regular part in the section in which we are discussing only on the boundary fluid.
This is introduced in \eqref{T_hat} and the boundary stress-energy tensor is given by
the coefficients of $\mathcal O(r^{-5})$ terms.

\item
$\widehat T^\mu{}_\nu$: the boundary stress-energy tensor,
or its regular part in the section in which we are discussing only the boundary fluid.
This is defined in \eqref{T_hat}.

\item
$T^\mu{}_\nu$: It is defined by \eqref{PhysSE},
$\widehat T^\mu{}_\nu = T^\mu{}_\nu + J^\mu \mathcal A_\nu - \delta^\mu{}_\nu J^\rho \mathcal A_\rho$.
This definition is used to define Milne-boost-invariants for a general background of $\mathcal A$. It is both Milne-boost invariant and gauge invariant.

\item
$\bar T^\mu{}_\nu$: defined by \eqref{Tbar},
$\bar T^\mu{}_\nu = \widehat T^\mu{}_\nu - J^\mu \widetilde A_\nu + \delta^\mu{}_\nu J^\rho \mathcal A_\rho$.
$\bar T^\mu{}_\nu - J^\mu \widetilde{\mathcal A}_\nu$
gives the standard non-relativistic fluids' stress-energy tensor. It
consists of the physical energy vector, physical momentum density, and physical stress tensor.
It is gauge invariant but not Milne-boost invariant..

\item
$J^\mu$: the current without counter terms defined in \eqref{Jdef}. It is regular even without the counter terms.
It corresponds to the mass current in a non-relativistic theory.

\item
$\widehat{\mathcal E}^\mu$: defined by $-\widehat T^\mu{}_\nu \hat v^\nu$ in \eqref{Edef}.
It corresponds to the (Milne boost invariant) energy vector.
It is not gauge invariant.

\item
$\widehat{\mathcal P}_\mu$: defined by $\widehat T^\rho{}_\nu \tau_\rho \hat e_a^\nu \hat e^a_\mu$ in \eqref{Pdef}.
It corresponds to the (Milne boost-invariant) momentum density.
It is different from the physical momentum density $\bar{\mathcal P}_\mu$,
which is not invariant under the Milne boost.
It is not gauge invariant.

\item
$\widehat{\mathcal T}^\mu{}_\nu$: defined by
$\widehat T^\rho_\sigma (\hat e^a_\rho \hat e_a^\mu)(\hat e_b^\sigma \hat e^b_\nu)$ in \eqref{Tdef}.
It corresponds to the (Milne boost invariant) stress tensor.
It is not gauge invariant.

\item
$\mathcal E^\mu$: defined by $- T^\mu{}_\nu \hat v^\nu$ in \eqref{Edef1}.
It corresponds to the (Milne boost invariant) energy vector.
It is gauge invariant.

\item
$\mathcal P_\mu$: defined by $T^\rho{}_\nu \tau_\rho \hat e_a^\nu \hat e^a_\mu$ in \eqref{Pdef1}.
It corresponds to the (Milne boost-invariant) momentum density.
It is different from the physical momentum density $\bar{\mathcal P}_\mu$,
which is not invariant under the Milne boost.
It is gauge invariant.

\item
$\mathcal T^\mu{}_\nu$: defined by
$T^\rho_\sigma (\hat e^a_\rho \hat e_a^\mu)(\hat e_b^\sigma \hat e^b_\nu)$ in \eqref{Tdef1}.
It corresponds to the (Milne boost invariant) stress tensor.
It is gauge invariant.

\item
$\bar{\mathcal E}^\mu$: defined by $-\bar T^\mu{}_\nu \bar v^\nu$ in \eqref{PhysE}.
It corresponds to the physical energy vector, which contains a contribution from the mass density.

\item
$\bar{\mathcal P}_\mu$: defined by $\bar T^\rho{}_\nu \tau_\rho \bar e_a^\nu \bar e^a_\mu$ in \eqref{PhysP}.
It corresponds to the physical momentum density, which contains a contribution from the mass density.

\item
$\bar{\mathcal T}^\mu{}_\nu$: defined by
$\bar T^\rho_\sigma (\bar e^a_\rho \bar e_a^\mu)(\bar e_b^\sigma \bar e^b_\nu)$ in \eqref{PhysT}.
It corresponds to the physical stress tensor, which contains  a contribution from the mass density.

\item
$\mathcal E$: (Milne boost invariant) energy density, or equivalently, internal energy density.
It is introduced in \eqref{FluidVariables} and equals $\mathcal E^0$.

\item
$P$: The pressure. It can be read off from the stress tensor and introduced in \eqref{FluidVariables}.

\item
$n$: it is defined by $1/a$.
It is introduced in \eqref{FluidVariables}.
It corresponds to the particle number density, or equivalently, the mass density.

\item
$T$: the temperature. It can be calculated as the Hawking temperature of the black hole \eqref{HawkingT}.

\item
$J^\mu_S$: the entropy current, which is defined from the volume form in \eqref{EntropyCurrent}
on the time-slice at the horizon.

\item
$s$: the entropy density, which is defined in \eqref{EntropyDensity}. It equals to $J^0_S$ in the non-relativistic case.

\item
$\eta$: the shear viscosity. It is introduced in \eqref{EtaInt} and can be read off from the stress-energy tensor, or the fluid equations.

\item
$\kappa$: the heat conductivity. It is introduced in \eqref{KappaInt}, and can be read off from the stress-energy tensor, or fluid equations.

\item
$\widetilde\Phi$: The Newton potential, which is $\widetilde \Phi = \widetilde{\mathcal A}_t$.
It is introduced around \eqref{NavierStokesNewton}.

\end{itemize}

\begin{table}
\renewcommand{\arraystretch}{1.2}
\begin{center}
\begin{tabular}{|l|c|c|c|c|c|}
\hline
This paper&
$\mathcal E^\mu$ &
$\mathcal E$ &
$\bar{\mathcal E}^\mu$ &
$\bar{\mathcal P}_\mu$ &
$\bar{\mathcal T}^{\mu\nu}$
\\\hline
Jensen \cite{Jensen:2014ama} &
$\widetilde{\mathcal E}^\mu$ &
$\varepsilon$ &
$\mathcal E^\mu$ &
$\mathcal P_\mu$ &
$T^{\mu\nu}$
\\\hline
\end{tabular}
\caption{
Correspondence of the fluid variables in this paper and those in \cite{Jensen:2014ama}.
In \cite{Jensen:2014ama}, the fluid variables do not contain contributions
from the external source.
}\label{table2}
\end{center}
\renewcommand{\arraystretch}{1}
\end{table}

\bigskip
\centerline{\bf Basic fields and constants on the gravitational  side}
\bigskip

\begin{itemize}
\item
$g_{\mu\nu}$: the metric. It is introduced in \eqref{BulkAction}.

\item
$A$: the gauge field. It is introduced in \eqref{BulkAction}.

\item
$\phi$: the dilaton. It is introduced in \eqref{BulkAction}.

\item
$G$: Newton's constant. It is introduced in \eqref{BulkAction}.

\item
$\hat A$: the gauge field with local Lorentz indices.  It is defined in \eqref{Ahat}.

\item
$\hat J$: the current with local Lorentz indices. It is defined in \eqref{VariationOfAction}.

\item
$h_{\mu\nu}$: the correction terms for the metric. It is introduced in \eqref{Defh}.

\item
$a_\mu$: the correction terms for the gauge field. It is introduced in \eqref{DefCorA}.

\item
$\varphi$: the correction terms for the dilaton. It is introduced in \eqref{DefCorPhi}.

\item
$R_{\mu\nu}$: The Ricci tensor. It appears first in \eqref{BulkAction}.

\item
$\Lambda$: the cosmological constant. It appears first in \eqref{BulkAction}.

\item
$T_{\mu\nu}^\text{(bulk)}$: the energy-momentum tensor in the bulk. It appears in \eqref{ConstEin}.

\item
$T^{\mu\nu}_\text{(BY)}$: The Brown-York tensor which is defined by \eqref{BYtensor},
$T^{\mu\nu}_\text{(BY)} = \frac{1}{8\pi G}\left(\gamma^{\mu\nu} K - K^{\mu\nu}\right)$.

\item
$n^\mu$: the normal vector to the boundary or horizon. It appears first in \eqref{Current}.

\item
$K_{\mu\nu}$: the extrinsic curvature on the boundary. It appears in \eqref{BYtensor}.
To be precise, it is defined on constant but finite $r$ surface.

\item
$\gamma_{\mu\nu}$: the induced metric on the boundary. It is introduced in \eqref{Indm}.
To be precise, it is defined on constant but finite $r$ surface.

\end{itemize}



\section{Calculation of the equations of motion at first order with $z=2$}\label{app:solution}

In this section, we calculate the first order solution
in the derivative expansion around \eqref{BGmetric}-\eqref{BGscalar}.
The coordinates can be chosen such that $v^i(x)=0$ at any given point,
therefore we may take $v^i(0)=0$ without loss of generality.
Although we will work in $v^i(0)=0$ coordinates,
the solution for $v^i(0)\neq 0$ can be obtained by boosting the solution uniformly.
{}From now on we work  at the point $x^\mu=0$, but we omit the subscript ``$(0)$'', hereafter.
Here, we take the following gauge conditions;
\begin{align}
 g_{rr}&=0 \ , &
 g_{r\mu} &\propto \hat v_\mu \ , &
 \tr[\bar g^{-1} h] &= 0 \ , &
 a_r &= 0 \ ,
\end{align}
where $\hat v^\mu = (1,v^i)$.


The correction terms can be classified by using the $SO(3)$ symmetry along the spatial directions.
The equations of motion are separated into that for scalar (sound mode) , vector mode and tensor mode. The
 sound mode consists of the following components
\begin{align}
 h_{tt} && h_{tr} && h^i{}_i && a_t && \varphi \ ,
\end{align}
and vector mode
\begin{align}
 h_{ti} && a_i \ ,
\end{align}
and the tensor mode is the traceless part of the metric
\begin{equation}
 h_{ij} \ .
\end{equation}

To simplify  the differential equations,
we redefine the correction terms for the metric
\begin{align*}
 h_{tt} && h_{tr} && h_{ti} && h_{xx} && h_{ij}
\end{align*}
 as follows
\begin{align}
 g_{tt} &= r^4(- f + h_{tt}) ,
\\
 g_{tr} &= \frac{r}{2} h_{tr} ,
\\
 g_{ti} &= r^2 h_{ti}
\\
 g_{ij} &= r^2 (\delta_{ij} + h_{xx}\delta_{ij} + h_{ij})
\end{align}
where $h_{xx}$ is the trace part and $h_{ij}$ is the traceless part in $x^i$-components.
The gauge condition gives additional constraints;
\begin{equation}
 2 h_{tr} + 3 h_{xx} = 0
\end{equation}
and the $(r,r)$- and $(r,i)$-components of the correction terms must vanish.
We define
\begin{equation}
 h_1 = \frac{1}{2} h_{xx}
\end{equation}
The other correction terms $a_\mu$ and $\varphi$ are similar to the definitions
\eqref{DefCorA} and \eqref{DefCorPhi}, but $a_r$ is eliminated by the gauge condition.

\subsection{The sound mode}

Some components of the equations of motion do not become
differential equations for the correction terms, but
give the constraints on the parameters.
{}From the Einstein equation we obtain
\begin{equation}
 n^\mu\gamma^{\nu\rho} R_{\mu\nu} = 8\pi G n^\mu\gamma^{\nu\rho} T_{\mu\nu}^\text{(bulk)} \label{ConstEin}
\end{equation}
where $n^\mu$ is the normal vector and $\gamma_{\mu\nu}$ is
the induced metric on $r=$const.\ surfaces.
In fact the above equation contains no correction terms.
For the sound mode, by contracting the \eqref{ConstEin} with $\hat v^\mu = (1,v^i)$,
we obtain the following constraint;
\begin{equation}
 0 = \frac{r_0^4}{2 r^4}\left(3 a r_0 \partial_i v^i +15 a \partial_t r_0 + 2 r_0 \partial_t p\right)
 + \left(a \partial_i v^i - \partial_t p\right)r \ ,
\label{2}\end{equation}
where
\begin{equation}
 p = \left(\frac{z+d-1}{2(z-1)} \mu \right)^{-1/2} \ .
\end{equation}
Equation (\ref{2}) must be satisfied for arbitrary $r$, and hence
the first and second terms must vanish independently,
\be
3 a r_0 \partial_i v^i +15 a \partial_t r_0 + 2 r_0 \partial_t p=0\;,
\label{4}\ee
\be
a \partial_i v^i - \partial_t p=0\;.
\label{5}\ee

{}From \eqref{MaxwellEq}, the $r$-component does not contain correction terms
and gives another constraint;
\begin{equation}
 - a \partial_i v^i + 2 \partial_t p - \partial_t a=0 \ .
\label{3}\end{equation}

After substituting the above constraints to the equations of motions for the sound mode,
we obtain the following differential equations;
\begin{align}
 0
 &=
 \frac{2 \sqrt{6} a_t'(r)}{a}
 -6 \sqrt{6} r^5 h_1'(r)+6 \sqrt{6} r_0^5 h_1'(r)+\sqrt{6} r^5
   h_{tt}'(r)+5 \sqrt{6} r^4 h_{tt}(r)
\notag\\
&\quad
 +r^6 \varphi''(r)-r r_0^5
   \varphi''(r)+6 r^5 \varphi'(r)-r_0^5 \varphi'(r)+30 r^4
   \varphi(r)  - 2 \sqrt{6} r^2 \partial_i v^i
\label{eqp}
\\
 0 &=
  r a_t''(r)-4 a_t'(r)+30 a r^5 h_1'(r)+5 \sqrt{6} a_0 r^5 \varphi'(r) \label{eqt}
\\
 0 &=
 - 12 a r^2 \partial_i v^i +8 a_t'(r)-72 a r^5 h_1'(r)-18
   a r_0^5 h_1'(r)-240 a r^4 h_1(r)
\notag\\
&\quad
+3 a r^6
   h_{tt}''(r)+30 a r^5 h_{tt}'(r)+60 a r^4 h_{tt}(r)+20
   \sqrt{6} a r^4 \varphi(r) \label{eqtt}
\\
 0 &=
 3 a \left(- 4 r^2 \partial_i v^i-6 \left(4 r^5+r_0^5\right)
   h_1'(r)+r^6 h_{tt}''(r)+10 r^5 h_{tt}'(r)\right)
\notag\\
&\quad
 +8 a_t'(r)+20
   \sqrt{6} a r^4 \varphi(r)
 -240 a r^4 h_1(r)+ 60 a r^4 h_{tt}(r) \label{eqtr}
\\
 0 &=
 - 12 h_1' + \sqrt{6} \varphi' - 3r h_1'' \label{eqrr}
\\
 0&=
-\frac{2 a_t'(r)}{a r^2}-3 r^4
   h_1''(r)+\frac{3 r_0^5 h_1''(r)}{r}-36 r^3
   h_1'(r)+\frac{21 r_0^5 h_1'(r)}{r^2}-120 r^2 h_1(r)
\notag\\
&\quad
 +3 r^3
   h_{tt}'(r)+15 r^2 h_{tt}(r)-5 \sqrt{6} r^2 \varphi(r) \label{eqxx}
  - 6 \partial_i v^i
\end{align}
where the first equation originates from \eqref{DilatonEOM} and
the second from the $t$-component of \eqref{MaxwellEq}.
The third to fifth equations are the $(t,t)$-, $(t,r)$- and $(r,r)$-components
of \eqref{EinsteinEq}. The last equation is the trace part of
the spatial component of \eqref{EinsteinEq}.
The above equations are not independent but
an appropriate combination gives the constraints,
which we have already imposed, and hence becomes trivial.

We first impose the constraints to the parameters and then
solve the differential equations. The solution for the sound modes is
\begin{align}
 h_{tt}
 &=
 \frac{2\sqrt 6}{5}\left(1 - \frac{r_0^5}{r^5}\right)\phi^{(0)} + \frac{2}{3r^2} \partial_i v^i
 - \frac{1}{r^5} \left(4 +  \frac{r_0^5}{r^5}\right)h_1^{(0)}
\notag\\
&\quad
 + \left(1 - \frac{6 r_0^5}{r^5}\right)h_2(r) + \left(1 - \frac{r_0^5}{r^5}\right) h'_2(r)
\\
 h_1
 &=
 \frac{1}{5}\sqrt{\frac23} \phi^{(0)} + \frac{1}{r^5} h_1^{(0)} + h_2(r)
\\
 a_t
 &=
 a_t^{(0)}
 - a \left(15 h_1^{(0)} + \sqrt 6 r^5 \phi^{(0)} - 15 r^5 h_2(r)\right) - \frac{1}{3} a r^3 \partial_i v^i
\\
 \varphi
 &=
 \frac{8}{5} \phi^{(0)} - \frac{\sqrt 6}{r^5} h_1^{(0)}
 + 3\sqrt{\frac{3}{2}} h_2(r) + \sqrt{\frac{3}{2}} \, r h_2'(r)
\end{align}
where $\phi^{(0)}$, $h_{tt}^{(0)}$, $h_1^{(0)}$, and $a_t^{(0)}$ are integration constants
and $h_2$ is given by
\begin{align}
 h_2(r)
 &=
 \frac{(-1)^{4/5} C_1}{5 r_0^4}{
   _2F_1\left(\frac{1}{2}-\frac{\sqrt{\frac{37}{5}}}{2},\frac{1}{2}+\frac{\sqrt{\frac{37}{5}}}{2};2;\frac{r^5}{r_0^5}\right)}
\notag\\
&\quad
 +\frac{C_2}{5 r^4}
   G_{3,3}^{2,1}\left(\frac{r^5}{r_0^5} \biggl|
\begin{array}{ccc}
 \frac{4}{5}, & \frac{1}{10} \left(13-\sqrt{185}\right), & \frac{1}{10}
   \left(13+\sqrt{185}\right) \\
 \frac{4}{5},& \frac{4}{5}, & -\frac{1}{5} \\
\end{array}
\right) ,
\end{align}
where ${}_pF_q$ and $G_{p,q}^{m,n}$ are hypergeometric function and Meijer $G$-function,
respectively \cite{Gradsteyn},
and $C_1$ and $C_2$ are integration constants.

\subsection{Vector mode}

As for the sound modes, the spatial component of \eqref{ConstEin} gives a constraint
\begin{align}
 0 =
 \partial_t \mathcal A_i + \mathcal A_j \partial_i v^j + 5 a r_0^4 \partial_i r_0
 + (r^5-r_0^5) \partial_i (p - a) \ .
\end{align}
Since this constraint must be satisfied at arbitrary $r$,
the first and second terms must vanish independently,
\be
\partial_t \mathcal A_i + \mathcal A_j \partial_i v^j + 5 a r_0^4 \partial_i r_0=0\;,
\label{6}\ee
\be
\partial_i (p - a)=0 \ .
\label{7}\ee

Then, the equations of motion for the vector modes are
\begin{align}
 0 &=
 r \left(-5 r r_0^4 \left(5 a \partial_i r_0 +r_0
   \partial_i a\right)+\left(r^5-r_0^5\right) a_i''(r)+5 a
   r^7 h_{ti}'(r)\right)+\left(7 r_0^5-2 r^5\right) a_i'(r)
\\
 0 &=
r^2 \left(-2 \left(5 a r_0^4 \partial_i r_0+\partial_i a
   \left(r_0^5-r^5\right)\right)+a r^2 \left(r_0^5-r^5\right)
   h_{ti}''(r)+4 a r \left(r_0^5-r^5\right) h_{ti}'(r)\right)
\notag\\
&\quad
-2
   \left(r^5-r_0^5\right) a_i'(r)
\label{eqti}
\\
 0 &=
 2 a_i' + 4 a r^3 h_{ti}' + a r^4 h_{ti}'' - 2 r^2 \partial_i p ,
\label{eqri}
\end{align}
where the first equation is the $x^i$-component of \eqref{MaxwellEq} and
the others are the $(t,x^i)$- and $(r,x^i)$-components of \eqref{EinsteinEq}, respectively.

The solution for the vector modes is
\begin{align}
 h_{ti}
 &= \int \frac{dr}{r^4}
 \left(h_{ti}^{(0)}-\frac{2}{a} a_i
 + \frac{2 r^3}{3a} \partial_i a\right)
\\
 a_i
 &=
 \left(r^5 + \frac{3 r_0^5}{2}\right) \int dr\, a_1 (r) ,
\end{align}
where $h_{ti}^{(0)}$ and $a_{i}^{(0)}$ are integration constants.
The function $a_1$ is given by
\begin{equation}
 a_1 (r) =
 - \frac{1}{3(r^5 - r_0^5)(2 r^5 + 3 r_0^5)^2}
 \left(
 30 h_{ti}^{(0)} a r^4 (r^5-r_0^5) - 3 r^7 C_3 + r^2 (8 r^{10} + 18 r_0^{10}) \partial_i a
 \right)
\end{equation}
and is expanded around $r=r_0$ as
\begin{equation}
 a_1 = \frac{3 C_3 - 26 r_0^5 \partial_i a}{375 r_0^7 (r-r_0)} + \cdots .
\end{equation}
In order for the solution to be regular at $r=r_0$,
we we must take
\begin{equation}
 C_3 = \frac{26}{3} r_0^5 \partial_i a
\end{equation}
and for the rest  we obtain
\begin{align}
 h_{ti} &= - \frac{a_i^{(0)}}{a} \frac{r^5 - r_0^5}{r^3}
\\
 a_i &= \frac{a}{2} h_{ti}^{(0)} + a_i^{(0)}\left(r^5 + \frac{3}{2}r_0^5\right) + \frac13 r^3 \partial_i a
\end{align}

\subsection{Tensor mode}

The equation of motion for the tensor mode is given by
\begin{align}
 0 = 2 (-6 r^5 + r_0^5) h_{ij}' + 2 (- r^6 + r r_0^5) h_{ij}'' - 3 r^2 (\partial_j v^i+ \partial_i v^j)\;.
\end{align}
There are no constraints for the tensor mode.

The solution  is
\begin{equation}
 h_{ij}
 =
 - \sigma_{ij} \int \frac{r^2 dr}{r^5-r_0^5}
 + C_4 \int \frac{dr}{r(r^5-r_0^5)}
\end{equation}
where
\begin{equation}
 \sigma_{ij} = \left(\partial_i v^j + \partial_j v^i\right)
 - \frac{2}{3} \partial_k v^k \delta_{ij}
\end{equation}
Regularity at $r_0$ implies $C_4=r_0^3$.
We finally obtain
\begin{equation}
 h_{ij}
 =
 - \sigma_{ij} \int \frac{(r^3 -r_0^3)dr}{r(r^5-r_0^5)}
\end{equation}

\section{Calculation of the stress-energy tensor}\label{app:stress}

Here, we calculate the stress-energy tensor.
For the solution \eqref{SolGeom}, the Brown-York tensor is obtained as
\begin{align}
 8 \pi G T_\text{(BY)}{}^0{}_0
 &=
 3 - \frac{3}{2 r^5} r_0^5 + \frac{1}{2 r^7} v^i \partial_i r_0^5 + \mathcal O(r^{-10}) \ ,
\\
 8 \pi G T_\text{(BY)}{}^i{}_0
 &=
 - v^i + \frac{1}{2 r^5} \left(- 4 r_0^5 v^i + \partial_i r_0^5\right)
 + \mathcal O(r^{-7}) \ ,
\\
 8 \pi G T_\text{(BY)}{}^0{}_i
 &=
 - \frac{1}{2 r^7} \partial_i r_0^5 + \mathcal O(r^{-10})\ ,
\\
 8 \pi G T_\text{(BY)}{}^i{}_j
 &=
 4 \delta_{ij} + \frac{1}{2 r^5} r_0^5 \delta_{ij}
 - \frac{1}{2 r^5} r_0^3 \sigma_{ij} + \mathcal O(r^{-7}) \ ,
\end{align}
to the first order in the derivative expansion.

The current is given by
\begin{align}
 16 \pi G J^0
 &=
 \frac{2}{a r^5}
 + \frac{2}{3 a r^7} \partial_i v^i
 + \frac{1}{a r^{10}} r_0^5 + \mathcal O(r^{-11}) \ ,
\\
 16\pi G J^i
 &=
 \frac{2}{a r^5} v^i + \frac{1}{a r^{10}} r_0^5 v^i + \mathcal O(r^{-11}) \ .
\end{align}

{}From (\ref{8}) we can obtain  the non-renormalized part of $T_r^\mu{}_\nu$ as
\begin{align}
 8 \pi G T_r^\text{(nr)}{}^0{}_0
 &=
 4 + \frac{1}{r^5} \left(-2 r_0^5 - \frac{1}{a} v^i \mathcal A_i \right) + \mathcal O(r^{-6}) \ ,
\label{nrT00}
\\
 8 \pi G T_r^\text{(nr)}{}^i{}_0
 &=
 \frac{1}{r^5} \left(- \frac{5}{2} r_0^5 v^i + \frac{1}{2} \partial_i r_0^5 \right) + \mathcal O(r^{-6}) \ ,
\\
 8 \pi G T_r^\text{(nr)}{}^0{}_i
 &=
 \frac{1}{a r^5} \mathcal A_i + \mathcal O(r^{-6}) \ ,
\\
 8 \pi G T_r^\text{(nr)}{}^i{}_j
 &=
 4 \delta_{ij} + \frac{1}{r^5} \left(\frac{1}{2} r_0^5 \delta_{ij} - \frac{1}{2} r_0^3 \sigma_{ij}
 + \frac{1}{a} v^i \mathcal A_j \right)
 + \mathcal O(r^{-6}) \ .
\label{nrTij}
\end{align}
Since the volume form on the boundary behaves as
\begin{equation}
 \sqrt{-\gamma} \sim r^5 \ ,
\end{equation}
the terms of $\mathcal O(r^{-5})$ become finite.
Those of $\mathcal O(r^0)$ diverge at the boundary, $r\to\infty$, but
can be subtracted by introducing the boundary cosmological constant term.
We can further introduce a boundary counter term proportional to $A^2$;
\begin{equation}
 S_\text{ct} =
 \frac{1}{16 \pi G }\int d^4 x \sqrt{-\gamma}
 \left(
 - 8 + C + \frac{5}{2} C e^{\lambda\phi}\gamma^{\mu\nu} A_\mu A_\nu
 \right) \ ,
\label{CT}
\end{equation}
This induces a counter term for the stress-energy tensor
\begin{align}
 8 \pi G T_r^\text{(ct)}{}^\mu{}_\nu
 = - (4 - C) \delta^\mu{}_\nu + C \frac{1}{2 r^5} r_0^5 \delta^\mu{}_\nu + \mathcal O(r^{-6}) \ .
\end{align}
Subtracting the counterterm, the renormalized stress-energy tensor becomes
\begin{align}
 \widehat T^0{}_0
 &=
 \frac{1}{8\pi G} \left[\left(-2+\frac{C}{2}\right) r_0^5 - \frac{1}{a} v^i \mathcal A_i \right]
 \ ,
\label{T00C}
\\
 \widehat T^i{}_0
 &=
 \frac{1}{8\pi G} \left(- \frac{5}{2} r_0^5 v^i + \frac{1}{2} \partial_i r_0^5 - \frac{1}{a} v^i v^j \mathcal A_j
 + \frac{1}{2} r_0^3 \sigma_{ij} v^j  \right)
 \ ,
\label{Ti0C}
\\
 \widehat T^0{}_i
 &=
 \frac{1}{8\pi G} \frac{1}{a} \mathcal A_i \ ,
\label{T0iC}
\\
 \widehat T^i{}_j
 &=
 \frac{1}{8\pi G} \left[\frac{1}{2}(1+C) r_0^5 \delta_{ij} - \frac{1}{2} r_0^3 \sigma_{ij}
 + \frac{1}{a} v^i \mathcal A_j \right]
t \ .
\label{TijC}
\end{align}
{}From the equations above we can read  the energy density $\mathcal E$ and pressure $P$
\begin{align}
 \mathcal E &= \frac{4-C}{16\pi G} r_0^5 \ , &
 P &= \frac{1+C}{16\pi G} r_0^5 \ .
\end{align}

The coefficient $C$ in the counter term can be fixed by
the regularity condition for the operator dual to the dilaton scalar $\phi$.
The vacuum expectation value of the operator $\mathcal O_\phi$ is given by
\begin{equation}
 \mathcal O_\phi
 = \lim_{r\to\infty} r^5 \mathcal O_r
\end{equation}
where
\begin{equation}
 \mathcal O_r
 =
 \frac{1}{\sqrt{-\gamma}} \frac{\delta S_r}{\delta\phi}
 =
 n^\mu \nabla_\mu \phi
\end{equation}
For our first order solution, the non-renormalized expectation value is calculated as
\begin{equation}
 \mathcal O^\text{(nr)}_r
 =
 \frac{1}{16\pi G}\left(\sqrt{6}
 - \sqrt{\frac32} \frac{r_0^5}{r^5}\right) + \mathcal O(r^{-6})
\end{equation}
The regular term is at $\mathcal O(r^{-5})$
while the counter terms becomes
\begin{align}
 \mathcal O^\text{(ct)}_r &= \frac{1}{16\pi G}\frac{5}{2} C \lambda e^{\lambda\phi} \gamma^{\mu\nu} B_\mu B_\nu
 = \frac{1}{16\pi G}\left( - \sqrt{6} \, C + \sqrt{6} \, C \frac{r_0^5}{r^5} \right) + \mathcal O(r^{-6}) \;.
\end{align}
The renormalized expectation value is
\begin{equation}
 {\mathcal O}_r = \frac{1}{16\pi G} \left(\sqrt{6}\,(1-C)
 - \sqrt{\frac{3}{2}}\,(1-2C)\frac{r_0^5}{r^5}\right) + \mathcal O(r^{-6}) \ .
\end{equation}
To obtain a finite  $\mathcal O_\phi$,
we must  take $C=1$, as the first term is divergent at the boundary.
We obtain
\begin{align}
 \widehat T^0{}_0
 &=
 \frac{1}{8\pi G} \left(-\frac{3}{2} r_0^5 - \frac{1}{a} v^i \mathcal A_i \right)
  \ ,
\\
 \widehat T^i{}_0
 &=
 \frac{1}{8\pi G} \left(- \frac{5}{2} r_0^5 v^i + \frac{1}{2} \partial_i r_0^5 - \frac{1}{a} v^i v^j \mathcal A_j
 + \frac{1}{2} r_0^3 \sigma_{ij} v^j  \right)
 \ ,
\\
 \widehat T^0{}_i
 &=
 \frac{1}{8\pi G} \frac{1}{a} \mathcal A_i  \ ,
\\
 \widehat T^i{}_j
 &=
 \frac{1}{8\pi G} \left( r_0^5 \delta_{ij} - \frac{1}{2} r_0^3 \sigma_{ij}
 + \frac{1}{a} v^i \mathcal A_j \right)
 \ ,
\end{align}
and
\begin{equation}
 {\mathcal O}_\phi = \frac{1}{16\pi G}\sqrt{\frac{3}{2}}\, {r_0^5} \ .
\end{equation}
Moreover the renormalized stress tensor (\ref{T00})-(\ref{Tij}) satisfies
\begin{equation}
 z \mathcal E = (d-1) P \ .
\end{equation}

\section{More on counter terms}\label{app:generalcounter}

We can also consider higher order terms of $A_\mu$ for the counter terms.
Due to the constraint on $a$ and $\mu$,
$A_\mu$ always appears with the factor of $e^{\lambda\phi}/2$.
Then, general counter terms are expressed as
\begin{equation}
 S_\text{ct} =
 \int d^4 x \sqrt{-\gamma}
 \left[
 - 8 + \sum_\alpha\frac{c_\alpha}{\alpha}
 - \sum_\alpha\frac{c_\alpha}{\alpha} \left(-\frac52 e^{\lambda\phi}\gamma^{\mu\nu} A_\mu A_\nu\right)^\alpha
 \right] \ ,
\end{equation}
where the boundary cosmological constant term is fixed such that
the boundary stress-energy tensor becomes finite.
Then, the stress-energy tensor becomes
\begin{align}
 \widehat T^0{}_0
 &=
 \frac{1}{8\pi G} \left[\left(-2+\frac{1}{2}\sum_\alpha c_\alpha\right) r_0^5
 - \frac{1}{a} v^i \mathcal A_i \right]
 \ ,
\label{T00Ca}
\\
 \widehat T^i{}_0
 &=
 \frac{1}{8\pi G} \left(- \frac{5}{2} r_0^5 v^i + \frac{1}{2} \partial_i r_0^5 - \frac{1}{a} v^i v^j \mathcal A_j
 + \frac{1}{2} r_0^3 \sigma_{ij} v^j  \right)
 \ ,
\label{Ti0Ca}
\\
 \widehat T^0{}_i
 &=
 \frac{1}{8\pi G} \frac{1}{a} \mathcal A_i \ ,
\label{T0iCa}
\\
 \widehat T^i{}_j
 &=
 \frac{1}{8\pi G} \left[\frac{1}{2}\left(1+\sum_\alpha c_\alpha\right) r_0^5 \delta_{ij} - \frac{1}{2} r_0^3 \sigma_{ij}
 + \frac{1}{a} v^i \mathcal A_j \right]
 \ .
\label{TijCa}
\end{align}

The dual operator to the dilaton is calculated from
\begin{equation}
 {\mathcal O}_r = \frac{1}{16\pi G} \left[\sqrt{6}\,\left(1-\sum_\alpha c_\alpha\right)
 - \sqrt{\frac{3}{2}}\,\left(1-\sum_\alpha \alpha c_\alpha\right)\frac{r_0^5}{r^5}\right]
 + \mathcal O(r^{-6}) \ .
\end{equation}
In order to regularize $\mathcal O_\phi = \lim_{r\to\infty} r^5\mathcal O_r$,
the coefficient $c_\alpha$ must satisfy
\begin{equation}
 \sum_\alpha c_\alpha = 1 \ .
\end{equation}
Then, the stress-energy tensor is the same as in \eqref{T00}-\eqref{Tij}.

\section{Regularity conditions of the gauge field at the horizon}\label{app:RegA}

If the guage field $A_\mu$ has non-zero $A_t$ at the horizon,
it becomes singular at the horizon.
It can be easily seen by making a  Wick rotation and
by considering the Polyakov loop wrapping on the time circle.
Although the horizon is a point in the imaginary (euclidean) time,
it becomes two surfaces, future and past horizon in real time.
In this section, we show that the singularity appears only
in the past horizon even for $A_t\neq 0$, if we use the Eddington-Finkelstein coordinates.
Here, we focus on the near-horizon region and discuss
the regularity of the gauge field at the horizon.

In the near horizon region, the metric of the non-extremal black holes
is universally given by the Rindler space;
\begin{equation}
 ds^2 = - r dt^2 + \frac{dr^2}{r} + (dx^i)^2 \ ,
\end{equation}
where $r=0$ is the horizon of the black hole.
The coordinates used above cover only the region  outside the horizon.
 
In order to change to  Eddington-Finkelstein coordinates,
we define null coordinates as
\begin{equation}
 t_\pm = t \pm \log r
\end{equation}
and then, the metric is expressed as
\begin{equation}
 ds^2 = - r dt_\pm^2 \pm 2 dr dt_\pm + (dx^i)^2 \ .
\end{equation}
The ingoing (outgoing) Eddington-Finkelstein coordinates (with $t_+$ ($t_-$))
also cover the region inside of the future (past) horizon.
The Kruskal coordinate is defined by
\begin{equation}
 x_\pm = e^{\pm t_\pm/2}
\end{equation}
and the metric becomes
\begin{equation}
 ds^2 = dx_+ dx_- + (dx^i)^2 \ .
\end{equation}
This covers all regions, and $x_+=0$ and $x_-=0$ are the past and future horizon, respectively.

If the gauge field has non-vanishing $A_t$ and regular $A_r$ at the horizon,
it is singular there. This can be seen as follows.
In the Kruskal coordinates, the gauge field becomes
\begin{equation}
 A = A_t dt = A_t \left(\frac{dx_+}{x_+} - \frac{dx_-}{x_-}\right) \ .
\end{equation}
Therefore, the gauge field is singular at the future and past horizon.
In the Eddington-Finkelstein coordinates, it is expressed as
\begin{equation}
 A = A_t \left(dt_\pm \mp \frac{dr}{r}\right) \ .
\end{equation}

However, if we take the ingoing Eddington-Finkelstein coordinates,
and if $A_r$ is not singular at the horizon,
the gauge field is singular only at the past horizon and
is regular at the future horizon.
In  Kruskal coordinates, the gauge field is expressed as
\begin{equation}
 A = A_+ dt_+ + A_r dr = A_+ \frac{dx_+}{x_+} + A_r \left(x_+ dx_- + x_- dx_+\right) \ .
\end{equation}
Therefore the gauge field is singular at the past horizon $x_+=0$ but regular
at the future horizon $x_-=0$.

\section{First order solution for general $z$}\label{app:general}

The correction terms can be calculated in a similar fashion to the $z=2$ case.
We define $h_{\mu\nu}$, $a_\mu$ and $\varphi$ as
\begin{align}
 g_{tt} &= r^{2z}(- f + h_{tt}) \ ,
\\
 g_{tr} &= \frac{r^{z-1}}{2} h_{tr} \ ,
\\
 g_{ti} &= r^2 h_{ti} \ ,
\\
 g_{ij} &= r^2 (\delta_{ij} + h_{xx}\delta_{ij} + h_{ij}) \ ,
\end{align}
\begin{equation}
 A = \left[a(x) \left(r^{5} - r_0^5(x)\right) - \mathcal A_i(x) v^i(x) \right]dt
 - a(x) r^2 dr + \mathcal A_i(x) dx^i + a_t dt + a_i dx^i \ ,
\end{equation}
and
\begin{equation}
 \varphi = \phi - \frac{1}{\lambda} \log(\mu r^{-6}) \ .
\end{equation}

In the $v^i(0)=0$ gauge, the constrains are expressed as
\begin{align}
 p(x) &= a(x) ,
\\
 \partial_t a &= a \partial_i v^i  \ ,
\\
 \partial_t r_0 &= - \frac13 r_0 \partial_i v^i \ ,
\\
 \partial_t A_i + A_j \partial_i v^j &= - \frac{z (z+3)}{2 (z-1)} r_0^{z+2} a \partial_i r_0 \ .
\end{align}
The first order solution for the sound modes is
\begin{align}
 h_{tt}
 &=
 2 \sqrt{6} \frac{\sqrt{z-1}}{z+3} \left(1-\frac{r_0^{z+3}}{r^{z+3}}\right)\phi^{(0)}
 - r^{-z-3} \left[2z -(z-3)\frac{r_0^{z+3}}{r^{z+3}}\right] h_1^{(0)}
\notag\\&\quad
 + r^{-z-3} h_{tt}^{(0)}
 + 4 \frac{z-1}{(z+3)^2} \left(1-\frac{r_0^{z+3}}{r^{z+3}}\right) \frac{a_t^{(1)}}{a}
\notag\\&\quad
 + r^{-z-2} \left(1-\frac{r_0^{z+3}}{r^{z+3}}\right) h_2(r)
 - r^{-z-3} \left[2z -(z-3)\frac{r_0^{z+3}}{r^{z+3}}\right] \int dr \, h_2(r)
\notag\\&\quad
 + \frac{2}{3} r^{-z} \partial_i v^i \ ,
\\
 h_1
 &=
 \sqrt{\frac23} \frac{\sqrt{z-1}}{z+3} \phi^{(0)}
 + \frac{2}{3} \frac{z-1}{(z+3)^2} \frac{a_t^{(1)}}{a}
 + r^{-z-3} h_1^{(0)}
 + r^{-z-3} \int dr \, h_2(r)
\\
 a_t
 &=
 a_t^{(0)} - 3 \frac{z+3}{z-1} a h_1^{(0)}
 - \frac{1}{z+3} r^{z+3} a_t^{(1)} + \sqrt{\frac{6}{z-1}} a r^{z+3} \phi^{(0)}
 + 3 \frac{z+3}{z-1} a \int dr \, h_2(r)
\notag\\&\quad
 - \frac{1}{3} a r^3 \partial_i v^i
\\
 \varphi
 &=
 \phi^{(0)} - \sqrt{6(z-1)} r^{-z-3} h_1^{(0)}
 - \sqrt{\frac32} \frac{(z-5)\sqrt{z-1}}{(z+3)^2} \frac{a_t^{(1)}}{a}
\notag\\&\quad
 + \sqrt{\frac32} \frac{1}{\sqrt{z-1}} r^{-z-3} \left(r h_3(r) - 2 (z-1) \int dr \, h_2(r) \right)
\end{align}
where $\phi^{(0)}$, $h_{tt}^{(0)}$, $h_1^{(0)}$, $a_t^{(0)}$ and $a_t^{(1)}$ are integration constants.
The function $h_2$ is the solution of the following differential equation;
\begin{align}
 0
 &=
 r^2 \left(r^{z+3} - r_0^{z+3} \right) h_2''(r)
 - r \left[z r^{z+3} + (2z+3) r_0^{z+3}\right] h_2'(r)
\notag\\&\quad
 - (z+2)\left[(2z+7) r^{z+3} + (z+2) r_0^{z+3} \right] h_2(r)
\end{align}
and is given by
\begin{align}
 h_2(r)
 &=
 C_1 r^{\frac12 \left(z+1+\sqrt{(z+3)(9z+19)}\right)} {}_2F_1
 \left(\alpha_-, \alpha_- ; 2 \alpha_- ; \frac{r_0^{z+3}}{r^{z+3}} \right)
\notag\\&\quad
 +
 C_2 r^{\frac12 \left(z+1-\sqrt{(z+3)(9z+19)}\right)} {}_2F_1
 \left(\alpha_+ , \alpha_+ ; 2 \alpha_+ ; \frac{r_0^{z+3}}{r^{z+3}} \right)
\end{align}
where $C_1$ and $C_2$ are integration constants, and
\begin{equation}
 \alpha_\pm = \frac{1}{2}\frac{\sqrt{z+3}\pm\sqrt{9z+19}}{z+3} \ .
\end{equation}

The first order solution for the vector modes is
\begin{align}
 h_{ti}
 &=
 \int \frac{dr}{r^{6-z}}
 \left(h_{ti}^{(0)}-\frac{2(z-1)}{a} a_i\right)
\\
 a_i
 &=
 \left( 2(z-1) r^{z+3} - (z-5) r_0^{z+3}\right) \int dr\, a_1 (r) ,
\end{align}
where $h_{ti}^{(0)}$ and $a_{i}^{(0)}$ are integration constants.
The function $a_1$ is given by
\begin{align}
 a_1 (r) &=
 \frac{r^7 a_2(r)}{(r^{z+3}-r_0^{z+3})[2(z-1) r^{z+3} - (z-5) r_0^{z+3}]^2}
%
\\
 a_2(r)
 &=
 C_3
 - (z+3) a r^{z-5} \left(r^{z+3} - r_0^{z+3}\right) h_{ti}^{(0)}
\notag\\&\quad
 + \frac{z+3}{2(z-1)} \frac{r_0^{z+2}}{r^5} a
 \left(10 (z-1) r^{z+3} - (z-5)(z-2) r^{z+3} \right) \partial_i r_0
\end{align}
and $a_2(r)$ is expanded around $r=r_0$ as
\begin{equation}
 a_2(r) = C_3 + \frac{z(z+3)^2 a r_0^{2z} \partial_i r_0}{2(z-1)} + \mathcal O(r-r_0) .
\end{equation}
In order for the solution to be non-singular at $r=r_0$,
we have to take
\begin{equation}
 C_3 = -\frac{z(z+3)^2 a r_0^{2z} \partial_i r_0}{2(z-1)} \ .
\end{equation}
and then, $a_2(r)$ becomes
\begin{align}
 a_2(r)
 &=
 - (z+3) a r^{z-5} \left(r^{z+3} - r_0^{z+3}\right) h_{ti}^{(0)}
\notag\\&\quad
 + \frac{z+3}{2(z-1)} \frac{r_0^{z+2}}{r^5} a
 \left(10 (z-1) r^{z+3} r_0^2 + z(z+3) r^5 r_0^z - (z-5)(z-2) r^{z+3} \right) \partial_i r_0 \ .
\end{align}

The first order solution for tensor modes is
\begin{equation}
 h_{ij}
 =
 - \sigma_{ij} \int \frac{r^2 dr}{r^{z+3}-r_0^{z+3}}
 + C_4 \int \frac{dr}{r(r^{z+3}-r_0^{z+3})} \ .
\end{equation}
The regularity at $r_0$ implies $C_4=r_0^3$.
Then, we obtain
\begin{equation}
 h_{ij}
 =
 - \sigma_{ij} \int \frac{(r^3 -r_0^3)dr}{r(r^{z+3}-r_0^{z+3})}   \ .
\end{equation}

\section{Counter terms for general $z$}\label{app:counter}

In order to obtain regular stress-energy tensor, we introduce the counter terms;
\begin{equation}
 S_\text{ct} =
 \frac{1}{16\pi G} \int d^4 x \sqrt{-\gamma}
 \left[
 - (4+2z) + C + \frac{z+d-1}{2(z-1)} C \, e^{\lambda\phi} \gamma^{\mu\nu} A_\mu A_\nu
 \right] \ .
\end{equation}
Since for general $z$, the volume form on the boundary behaves as
\begin{equation}
 \sqrt{-\gamma} \sim r^{z+3} \ ,
\end{equation}
the regular contribution to the stress-energy tensor is given by
$\mathcal O(r^{-z-3})$ terms of $T_r^\mu{}_\nu$.
Then, the renormalized stress-energy tensor is obtained as
\begin{align}
 \widehat T^0{}_0
 &=
 \frac{1}{8\pi G}\left[\left(\frac{C}{2} - \frac{z+2}{2}\right) r_0^{z+3}
 - \frac{z-1}{a} v^i \mathcal A_i \right]
 \ ,
\\
 \widehat T^i{}_0
 &=
 \frac{1}{8\pi G} \left(- \frac{z+3}{2} r_0^{z+3} v^i
 + \frac{z(z+3)}{4(z-1)} r_0^{2z} \partial_i r_0
 - \frac{z-1}{a} v^i v^j A_j
 + \frac{1}{2} r_0^3 \sigma_{ij} v^j  \right)
  \ ,
\\
 \widehat T^0{}_i
 &=
 \frac{1}{8\pi G}\frac{z-1}{a} A_i \ ,
\\
 \widehat T^i{}_j
 &=
 \frac{1}{8\pi G} \left[\frac{1}{2} (1+C) r_0^{z+3} \delta_{ij}
 - \frac{1}{2} r_0^3 \sigma_{ij}
 + \frac{z-1}{a} v^i A_j \right]
 \ .
\end{align}
For $C=z-1$, the energy density and the pressure becomes
\begin{align}
 \mathcal E &= \frac{3}{16\pi G} r_0^{z+3} \ , &
 P &= \frac{z}{16\pi G} r_0^{z+3} \ ,
\end{align}
and they satisfy
\begin{equation}
 z \mathcal E = (d-1) P  \ .
\end{equation}

The constant $C$ in the counter terms can be fixed by the regularity of
the dual operator to the dilaton $\phi$.
By introducing the counter terms, the expectation value of the operator becomes
\begin{align}
 {\mathcal O}_\phi
 &=
 \lim_{r\to\infty} r^{z+3}\,\mathcal O_r
\\
 {\mathcal O}_r
 &=
 \frac{1}{16 \pi G}\left[\sqrt{6(z-1)} - C \sqrt{\frac{6}{z-1}}
 + \left(C \sqrt{\frac{6}{z-1}} - \sqrt{\frac{3(z-1)}{2}}\right) \frac{r_0^{z+3}}{r^{z+3}} \right]
\notag\\&\quad
 + \mathcal O(r^{-(z+4)}) \ .
\end{align}
In order for the above expression to be regular,
we have to take $C=z-1$, and then, we obtain
\begin{align}
 \widehat T^0{}_0
 &=
 \frac{1}{8\pi G} \left(- \frac{3}{2} r_0^{z+3} - \frac{z-1}{a} v^i A_i \right)
 \ ,
\\
 \widehat T^i{}_0
 &=
 \frac{1}{8\pi G} \left(-\frac{z+3}{2} r_0^{z+3} v^i + \frac{z(z+3)}{4(z-1)} r_0^{2z} \partial_i r_0
 - \frac{z-1}{a} v^i v^j A_j
 + \frac{1}{2} r_0^3 \sigma_{ij} v^j  \right)
 \ ,
\\
 \widehat T^0{}_i
 &=
 \frac{1}{8\pi G}\frac{z-1}{a} A_i  \ ,
\\
 \widehat T^i{}_j
 &=
 \frac{1}{8\pi G} \left(\frac{z}{2} r_0^{z+3} \delta_{ij} - \frac{1}{2} r_0^3 \sigma_{ij}
 + \frac{z-1}{a} v^i A_j \right)
 \ ,
\end{align}
and
\begin{equation}
 {\mathcal O}_\phi
 =
 \frac{1}{16\pi G}\sqrt{\frac{3(z-1)}{2}}\,{r_0^{z+3}} \ .
\end{equation}

\end{document}